\documentclass[aps,pre,twocolumn,groupedaddress,showpacs]{revtex4-2}
\newcommand{\bec}[1]{\mbox{\boldmath $ #1$}}
\usepackage{bm}
\usepackage{graphicx}
\usepackage{color}

{}
\newcommand{\meanUU}{\overline{\bm{U}}}

{}
{}

\newcommand{\meanU}{\overline{U}}

\begin{document}
\title{Energy and flux budget closure theory for passive scalar in stably stratified turbulence}
\author{N. Kleeorin$^{1,2}$}
%\email{nat@bgu.ac.il}
\author{I.~Rogachevskii$^{1,2}$}
\email{gary@bgu.ac.il}
%\homepage{http://www.bgu.ac.il/~gary}
\author{S. Zilitinkevich$^{3,4}$}
%\email{sergej.zilitinkevich@fmi.fi}

\bigskip
\affiliation{
$^1$Department of Mechanical Engineering, Ben-Gurion University of the Negev, Beer-Sheva
 8410530, Israel
 \\
$^2$Nordita, Stockholm University and KTH Royal Institute of Technology, 10691 Stockholm, Sweden
 \\
$^3$Institute for Atmospheric and Earth System Research (INAR), University of Helsinki, 00014 Helsinki, Finland
 \\
$^4$Finnish Meteorological Institute, 00101 Helsinki, Finland
}

\date{\today}
\begin{abstract}
The energy and flux budget (EFB) closure theory for a passive scalar (non-buoyant and non-inertial particles
or gaseous admixtures) is developed for stably stratified turbulence. The physical background of the EFB turbulence closures is based on the budget equations for the turbulent kinetic and potential energies and turbulent fluxes of momentum and buoyancy, as well as the turbulent flux of particles. The EFB turbulence closure is designed for stratified geophysical flows from neutral to very stable stratification and it implies that turbulence is maintained by the velocity shear at any stratification.
In a steady-state, expressions for the turbulent flux of passive scalar and the anisotropic non-symmetric turbulent diffusion tensor are derived, and universal flux Richardson number dependencies of the components of this tensor are obtained.
The diagonal component in the vertical direction of the turbulent diffusion tensor is suppressed by strong stratification, while the diagonal components in the horizontal directions are not suppressed,
and they are dominant in comparison with the other components of turbulent diffusion tensor. This implies that any initially created strongly inhomogeneous particle cloud is evolved into a thin pancake in horizontal plane with very slow increase of its thickness in the vertical direction.
The turbulent Schmidt number (the ratio of the eddy viscosity and the vertical turbulent diffusivity of passive scalar) increases linearly with the gradient Richardson number. The physics of such behaviour is related to the buoyancy force that causes a correlation between fluctuations of the potential temperature and the particle number density. This correlation that is proportional to the product of the vertical turbulent particle flux and the vertical gradient of the mean potential temperature, reduces the vertical turbulent particle flux. Considering the applications of these results to the atmospheric boundary-layer turbulence, the theoretical relationships are derived which allow to determine the turbulent diffusion tensor as a function of the vertical coordinate measured in the units of the local Obukhov length scale. The obtained relations are potentially useful in modelling applications of particle dispersion in the atmospheric boundary-layer turbulence and free atmosphere turbulence.
\end{abstract}

%e-print: NORDITA-2021-

\maketitle
\section{Introduction}

Turbulence and the associated turbulent transport of passive scalar
have been investigated systematically for more than a hundred years
in theoretical, experimental and numerical
studies. \cite{MY71,MY75,MC90,F95,P2000,LE08,DA13,RI21}
But some fundamental questions remain.
This is particularly true in applications such as geophysics and astrophysics,
where the governing parameter values are too
large to be modelled either experimentally or numerically.

The classical theory of atmospheric turbulence implies that the turbulent flux of any quantity is a product of a mean gradient of the quantity and a turbulent-exchange coefficient (e.g., eddy viscosity, eddy diffusivity, etc.). \cite{K41,K42,MY71,MY75} This corresponds to a down-gradient transport where the turbulent-exchange coefficients are proportional to density of the turbulent kinetic energy multiplied by turbulent timescale. This has been originally formulated for neutrally stratified turbulence. \cite{K41,K42}

Many turbulence closure models of stratified turbulence in meteorological applications \cite{MY71,MY75,UB05} have been based only on the density of the turbulent kinetic energy equation, not considering an evolution of the density of the turbulent potential energy proportional to the second moment of potential temperature fluctuations. In stable stratification, such turbulence closure models have resulted in the erroneous conclusion that shear-generated turbulence inevitably decays and that the flow becomes laminar in ''supercritical" stratifications (at gradient Richardson number exceeding some critical value), \cite{C61,M86} where the gradient Richardson number is the ratio of the squared Brunt-V\"{a}is\"{a}l\"{a} frequency (proportional to the gradient of the mean potential temperature) to the squared mean velocity shear.

Contradictions of this conclusion via the well-documented universal existance of turbulence in strongly ''supercritical'' conditions typical of the free atmosphere and the deep ocean, \cite{SF01,O01,BAN02,PAR02,MO02,LA04,SR10,M10,M14} have been attributed to some unknown mechanisms and, in practical applications, mastered heuristically. \cite{M10,M14,CAN09}
Numerous alternative turbulence closures in stratified turbulence have been formulated
using the budget equations for various turbulent parameters (in addition to the density of the turbulent
kinetic energy) together with heuristic hypotheses and empirical relationships. \cite{WT03,UB05}

As an alternative, the energy and flux budget (EFB) theory of turbulence closure for stably stratified dry atmospheric flows has been recently developed. \cite{ZKR07,ZKR08,ZKR09,ZKR10,ZKR13,KRZ19} In agreement with wide experimental evidence, the EFB theory shows that high-Reynolds-number turbulence is maintained by shear in any stratification, and the ''critical Richardson number", treated many years as a threshold between the turbulent and laminar regimes, actually separates two turbulent regimes: the strong turbulence typical of atmospheric boundary layers and the weak three-dimensional turbulence typical of the free atmosphere or deep ocean, and characterized by strong decrease in heat transfer in comparison to momentum transfer. The EFB theory have been verified against scarce data from the atmospheric experiments, direct numerical simulations (DNS), large-eddy simulations (LES) and laboratory experiments relevant to the steady state turbulence regime. \cite{ZKR07,ZKR08,ZKR13}
Following the EFB closure, other turbulent closure models
also do not imply a critical Richardson number. \cite{MSZ07,GAS07,CCH08,LPR08,SB08,S09,KC09,K10,LK16,L19}

In stably stratified turbulence, large-scale internal gravity waves result in additional vertical turbulent flux of momentum and additional productions of the densities of the turbulent kinetic energy (TKE), turbulent potential energy (TPE) and turbulent flux of potential temperature. \cite{ZKR09,KRZ19} For the stationary, homogeneous regime, the EFB theory in the absence of the large-scale internal gravity waves (IGW) yields universal dependencies of the flux Richardson number, the turbulent Prandtl number, the ratio of TKE to TPE, and the normalised vertical turbulent fluxes of momentum and heat on the gradient Richardson number. \cite{ZKR07,ZKR13} Due to the large-scale IGW, these dependencies lose their universality.  The maximal value of the flux Richardson number (universal constant  0.2-0.25 in the no-IGW regime) becomes strongly variable in the turbulence with large-scale IGW.
In the vertically homogeneous stratification, the flux Richardson number increases with increasing wave energy.
In addition, the large-scale internal gravity waves reduce anisotropy of turbulence. Predictions from this theory are consistent with available data from atmospheric and laboratory experiments, DNS and LES. \cite{ZKR09,KRZ19}

In the present study we develop the energy and flux budget turbulence closure theory for passive scalar
(non-buoyant, non-inertial particles and gaseous admixtures) for stably stratified turbulence.
We find that the vertical turbulent diffusion coefficient of passive scalar is strongly reduced for large gradient Richardson number, and turbulent Schmidt number (the ratio of the eddy viscosity and the vertical turbulent diffusivity of passive scalar) increases linearly with the gradient Richardson number.

For the atmospheric boundary-layer turbulence, we derive the theoretical relationships for the vertical
profiles of the turbulent diffusion tensor and the turbulent Schmidt number.
This study can be useful in modelling applications for the atmospheric boundary-layer turbulence and free atmosphere turbulence.
For example, transport of pollutants in the atmospheric turbulent flows is
an important environmental problem (see reviews \cite{TS13,J15,LM16} and references therein).
In stratified flows, the turbulent Schmidt number increases with the level of stratification. \cite{GA17,HS08} This is consistent with the observation that stratification acts more effectively against mass diffusivity than against momentum diffusivity. In spite of many studies, there is still controversy about the proper parameterization of the turbulent Schmidt number for the various environmental flows. \cite{GA17,KLL16}

This paper is organized as follows.
In Section II we outline the EFB theory for turbulence, where
we formulate governing equations for the energy and flux budget
turbulence-closure theory for stably stratified turbulence
and consider the steady-state and homogeneous regime of turbulence.
In Section III we develop the EFB theory for passive scalars,
deriving the budget equation for the turbulent flux of particles,
which yields the expression for turbulent diffusion tensor.
In Section IV we consider the applications of the obtained
results to the atmospheric boundary-layer turbulence and discuss in this section
the theoretical relationships potentially useful in modelling applications.
Finally, conclusions are drawn in Section V.
In Appendix~A we derive the budget equation for
the correlation function for fluctuations of particle number
density and temperature.

\section{The EFB theory for stably stratified turbulence}

In this study we consider fully developed stably stratified turbulence for geophysical flows
where typical vertical gradients of the mean velocity, potential temperature and other variables
are much larger than the horizontal gradients, so that direct effects of the mean-flow
horizontal gradients on turbulent statistics are negligible.
In such flows vertical scales of motions are much smaller than horizontal scales,
and the mean-flow vertical velocity is much smaller than the horizontal velocities.
This implies that the vertical turbulent transports are comparable with or even dominate
the mean flow vertical advection, whereas the stream-wise horizontal turbulent transport
is usually negligible compared to the horizontal advection.

In this section we formulate the energy and flux budget (EFB) closure theory
for stably stratified turbulence based on the budget equations for the densities of turbulent
kinetic and potential energies, and turbulent fluxes of momentum and heat.
In our analysis, we use budget equations for the one-point second moments
to develop a mean-field theory. We do not study small-scale structure
of turbulence (i.g., higher moments for turbulent quantities and intermittency).
In particular, we study  large-scale long-term dynamics, i.e.,
we consider effects in the
spatial scales which are much larger than the integral scale of turbulence
and in timescales which are much longer than the turbulent timescales.

\subsection{Budget equations for turbulence}
\label{sect-IIA}

In the framework of the energy and flux budget turbulence theory \cite{ZKR07,ZKR13}, we use
the budget equations for the density of turbulent kinetic energy (TKE) $E_{\rm K}=\langle {\bm u}^2 \rangle/2$, the intensity of potential temperature fluctuations $E_\theta=\langle \theta^2 \rangle/2$, the turbulent flux $F_i = \langle u_i \, \theta \rangle$ of potential temperature and the off-diagonal components of the Reynolds stress $\tau_{iz} =\langle u_i \, u_z \rangle$ with $i=x, y$:
\begin{eqnarray}
{DE_{\rm K} \over Dt} + \nabla_z \, \Phi_{\rm K} &=& - \tau_{i z} \, \nabla_z \meanU_i + \beta \, F_z - \varepsilon_{\rm K} ,
 \label{C1}
\end{eqnarray}
\begin{eqnarray}
{D E_\theta \over Dt} + \nabla_z \, \Phi_\theta &=& - F_z \, \nabla_z \overline{\Theta} - \varepsilon_\theta,
 \label{C2}
\end{eqnarray}
\begin{eqnarray}
{\partial F_i \over \partial t} + \nabla_z \, {\bm \Phi}_{i}^{({\rm F})}
&=& - \tau_{iz} \, \nabla_z \overline{\Theta} \, \delta_{i3} + 2 \beta \, E_\theta \, \delta_{i3} - {1 \over \rho_0} \, \langle \theta \, \nabla_i p \rangle
\nonumber\\
&& - F_z \, \nabla_z \meanU_i - \varepsilon_i^{({\rm F})} ,
\label{C3}
\end{eqnarray}
\begin{eqnarray}
{D \tau_{iz} \over Dt} + \nabla_z \, \Phi_{i}^{(\tau)} &=& - 2 E_z \, \nabla_z \meanU_i - \varepsilon_{i}^{(\tau)} ,
 \label{C15}
\end{eqnarray}
where $D / Dt = \partial /\partial t + \meanUU {\bf \cdot} \bec{\nabla}$,
the fluid velocity $\meanUU + {\bf u}$ is characterized by the mean fluid velocity $\meanUU(z)=(\meanU_x, \meanU_y, 0)$ and fluctuations ${\bf u}=(u_x, u_y, u_z)$,
$E_z=\langle u_z^2 \rangle/2$ is the density of the vertical turbulent kinetic energy,
$F_z=\langle u_z \, \theta \rangle$ is the vertical component of the turbulent heat flux,
$\delta_{ij}$ is the Kronecker unit tensor, the angular brackets imply ensemble averaging,
$\Theta = T (P_\ast / P)^{1-\gamma^{-1}}$ is the potential temperature,
$T$ and $P$ are the fluid temperature and pressure with their reference values, $T_\ast$  and $P_\ast$, respectively, $\gamma = c_{\rm p}/c_{\rm v}$ is the specific heat ratio, the potential temperature $\Theta = \overline{\Theta} + \theta$ is characterized by the mean potential temperature $\overline{\Theta}(z)$ and fluctuations $\theta$, the fluid pressure $P = \overline{P} + p$ is characterized by the mean pressure $\overline{P}$ and fluctuations $p$, $\, \beta=g/T_\ast$  is the buoyancy parameter, ${\bf g}$  is the gravity acceleration, and $\rho_0$ is the fluid density. We use here the Boussinesq approximation.

The term $\Pi_K \equiv - \tau_{i z} \, \nabla_z \meanU_i = K_{\rm M} S^2$ is the rate of energy production for the shear-produced turbulence, $K_{\rm M}$ is the turbulent viscosity, $S = \left[(\nabla_z \meanU_x)^2 + (\nabla_z \meanU_y)^2\right]^{1/2}$ is the large-scale velocity shear.
The terms $\Phi_{\rm K}$, $\Phi_\theta$, ${\bm \Phi}_{i}^{({\rm F})}$ and $\Phi_{i}^{(\tau)}$ include
the third-order moments. In particular, $\Phi_{\rm K} = \rho_0^{-1} \langle u_z \, p\rangle + (\langle u_z \, {\bf u}^2 \rangle - \nu \, \nabla_z \langle {\bf u}^2 \rangle)/2$ determines the flux of $E_{\rm K}$; $\, \Phi_\theta = \left(\langle u_z \, \theta^2 \rangle - \kappa \, \nabla_z \langle  \theta^2 \rangle\right)/2$ describes the flux of $E_\theta$; ${\bm \Phi}_{i}^{({\rm F})} = \langle u_i \, u_z \, \theta\rangle - \nu \, \langle \theta \, (\nabla_z u_i) \rangle - \kappa \, \langle u_i \, (\nabla_z \theta) \rangle$ determines the flux of $F_i$ and
$\Phi_{i}^{(\tau)}=\langle u_i \, u_z^2 \rangle + \rho_0^{-1} \, \langle p \, u_i\rangle - \nu \, \nabla_z \tau_{iz}$ describes the flux of $\tau_{iz}$.

The term $\varepsilon_{\rm K} = \nu \, \langle (\nabla_j u_i)^2 \rangle$ is the dissipation rate of the density of the turbulent kinetic energy, $\varepsilon_\theta = \kappa \, \langle ({\bm \nabla} \theta)^2 \rangle$ is the dissipation rate of the intensity of potential temperature fluctuations $E_\theta$ and $\varepsilon_i^{({\rm F})} = (\nu + \kappa) \, \langle (\nabla_j u_i) \, (\nabla_j \theta) \rangle$  is the dissipation rate of the turbulent heat flux $F_i$, where $\nu$ is the kinematic viscosity of fluid and $\kappa$ is the temperature diffusivity.
The term $\varepsilon_{i}^{(\tau)} = \varepsilon_{iz}^{(\tau)} - \beta \, F_i - Q_{iz}$ in Eq.~(\ref{C15}) is the ''effective dissipation rate"  of the off-diagonal components of the Reynolds stress $\tau_{iz}$, where $Q_{ij} = \rho_0^{-1} (\langle p \nabla_i u_j\rangle + \langle p \nabla_j u_i\rangle)$ and $\varepsilon_{iz}^{(\tau)}=2 \nu \, \langle (\nabla_j u_i) \, (\nabla_j u_z) \rangle$ is the molecular-viscosity dissipation rate (see below). \cite{ZKR07,ZKR13}

The first term, $- \tau_{iz} \, \nabla_z \overline{\Theta} \, \delta_{i3}$,  in the right hand side of Eq.~(\ref{C3}) contributes to the traditional vertical turbulent flux of potential temperature
which describes the classical gradient mechanism of the turbulent heat transfer.
On the other hand, the second and third terms in the right hand side
of Eq.~(\ref{C3}) describe a non-gradient contribution to the vertical turbulent flux of
potential temperature. In stably stratified flows
the gradient and non-gradient contributions to the vertical turbulent flux of
potential temperature have opposite signs [see Eq.~(\ref{X27})]. This implies that
the non-gradient contribution decreases the traditional gradient turbulent flux.

The budget equations for the components of the turbulent kinetic energies
$E_\alpha = \langle u_\alpha^2\rangle/2$ along the $x$, $y$
and $z$ directions can be written as follows:
\begin{eqnarray}
{DE_\alpha \over Dt} + \nabla_z \, \Phi_{\alpha} &=& - \tau_{\alpha z} \, \nabla_z \meanU_\alpha + \delta_{\alpha 3} \, \beta \, F_z  + {1 \over 2} Q_{\alpha\alpha}
- \varepsilon_{\alpha},
\nonumber\\
\label{C4}
\end{eqnarray}
where $\alpha=x,y,z$, the term $\varepsilon_{\alpha} = \nu \, \langle (\nabla_j u_\alpha)^2 \, \rangle$ is the dissipation rate of the turbulent kinetic energy components $E_\alpha$ and
$\Phi_{\alpha}$ determines the flux of $E_\alpha$. Here
$\Phi_{z}= \rho_0^{-1} \langle u_z \, p\rangle + (\langle u_z^3 \rangle - \nu \, \nabla_z \langle u_z^2 \rangle) / 2$ and $\Phi_{x,y}= (\langle u_z \, u_{x,y}^2 \rangle - \nu \, \nabla_z \langle u_{x,y}^2 \rangle) / 2$.
The terms $Q_{\alpha\alpha} = 2 \rho_0^{-1} \langle p \nabla_\alpha u_\alpha\rangle$ are the diagonal terms of the tensor $Q_{ij}$.
In Eq.~(\ref{C4}) we do not apply the summation convention for the double Greek indices.
Different aspects related to budget equations~(\ref{C1})--(\ref{C4})
have been discussed in a number of publications.
\cite{ZKR07,ZKR08,ZKR09,ZKR10,ZKR13,KRZ19,KF94,CM93,CCH02,OT87}

The density of turbulent potential energy (TPE) is determined by potential temperature fluctuations and is defined as $E_{\rm P} = (\beta^2 / N^2) \, E_\theta$,
where $N^2 = \beta \, \nabla_z \overline{\Theta}$, and $N$ is the Brunt-V\"{a}is\"{a}l\"{a} frequency.
The budget equation for the density of turbulent potential energy $E_{\rm P} = (\beta^2 / N^2) \, E_\theta$ reads:
\begin{eqnarray}
{\partial E_{\rm P} \over \partial t} + \nabla_z \, \Phi_{\rm P} = - \beta \, F_z - \varepsilon_{_{\rm P}} ,
\label{X24}
\end{eqnarray}
where $- \beta \, F_z$ is the rate of production of the turbulent potential energy density, $\Phi_{\rm P} = (\beta^2 / N^2) \, \Phi_\theta$ is the flux of $E_{\rm P}$ and
$\varepsilon_{_{\rm P}} = (\beta^2 / N^2) \, \varepsilon_\theta$ is the dissipation
rate of the density of the turbulent potential energy.
Using Eqs.~(\ref{C1}) and~(\ref{X24}), we obtain the budget equation for
the density of the total turbulent energy $E_{\rm T}=E_{\rm K}+E_{\rm P}$ as \cite{ZKR07,ZKR13}
\begin{eqnarray}
{\partial E_{\rm T} \over \partial t} + \nabla_z \, \Phi_{\rm T} = - \tau_{i z} \, \nabla_z \meanU_i  - \varepsilon_{_{\rm T}} ,
\label{X26}
\end{eqnarray}
where $\Phi_{\rm T}=\Phi_{\rm K}+\Phi_{\rm P}$ is the flux of $E_{\rm T}$ and $\varepsilon_{_{\rm T}}=\varepsilon_{_{\rm K}}+\varepsilon_{_{\rm P}}$
is the dissipation rate of the density of the total turbulent energy.

\subsection{Steady-state and homogeneous regime of turbulence}
\label{sect-IIB}

The discussed energy and flux budget turbulence closure theory
for stably stratified flows assumes the following:
\begin{itemize}
\item{The characteristic times of variations of the densities of
the turbulent kinetic energy (TKE) $E_{\rm K}$, the vertical and horizontal TKE $E_\alpha$,
the intensity of potential temperature fluctuations $E_\theta$
(and the turbulent potential energy $E_{\rm P}$),
the turbulent flux $F_i$ of potential temperature and
the turbulent flux  $\tau_{iz}$ of momentum
(i.e., the off-diagonal components of the Reynolds stress)
are much larger than the turbulent timescale.
This allows us to obtain steady-state solutions
of the budget equations~(\ref{C1})--(\ref{X26}) for TKE, TPE, TTE, $F_i$, $E_\alpha$ and $\tau_{iz}$
for a stably stratified turbulence.
}
\item{We neglect the divergence of the fluxes of TKE, TPE, $E_\alpha$, $F_i$ and $\tau_{iz}$
for a steady-state homogeneous regime of a stably stratified turbulence
(i.e., we neglect the divergence of third-order moments).
}
\item{Dissipation rates of TKE, TPE, $E_\alpha$ and $F_i$ are expressed using
the Kolmogorov  hypothesis, i.e.,
$\varepsilon_{\rm K}=E_{\rm K}/t_{\rm T}$, $\varepsilon_\theta=E_\theta/(C_{\rm p} \, t_{\rm T})$,
$\varepsilon_{\alpha\alpha}^{(\tau)} = E_\alpha/3t_{\rm T}$ and
$\varepsilon_i^{({\rm F})}=F_i/(C_{\rm F} \, t_{\rm T})$, where
$t_{\rm T}=\ell_0 /E_{\rm K}^{1/2}$ is the turbulent dissipation timescale, $\ell_0$ is the integral scale of turbulence, and $C_{\rm p}$  and $C_{\rm F}$ are dimensionless empirical constants. \cite{K41,K42,MY71,MY75,RI21}
}
\item{
The term $\varepsilon_{i}^{(\tau)} = \varepsilon_{iz}^{(\tau)} - \beta \, F_i - Q_{iz}$ in Eq.~(\ref{C15}) is the effective dissipation rate of the off-diagonal components of the Reynolds stress $\tau_{iz}$, where $\varepsilon_{iz}^{(\tau)}=2 \nu \, \langle (\nabla_j u_i)^2 \rangle$ is the molecular-viscosity dissipation rate of $\tau_{iz}$, that is small because the smallest eddies associated with viscous dissipation are presumably isotropic. \cite{LPR09} In the framework of EFB theory, the role of the dissipation of $\tau_{iz}$ is assumed to be played by the combination of terms $- \beta \, F_i - Q_{iz}$, and it is assumed that $\varepsilon_{i}^{(\tau)} =\tau_{iz}/(C_\tau \, t_{\rm T})$, where $C_\tau$ is the effective-dissipation time-scale empirical constant. \cite{ZKR07,ZKR13}
}
\item{We assume that the term $\rho_0^{-1} \, \langle \theta \,
\nabla_z p \rangle$ in Eq.~(\ref{C3}) for the vertical turbulent flux of potential
temperature is parameterised by $\tilde C_\theta \, \beta \, \langle \theta^2
\rangle$ with $\tilde C_\theta < 1$, where $\tilde C_\theta$  is dimensionless empirical constant.
This implies that $\beta \, \langle \theta^2
\rangle - \rho_0^{-1} \, \langle \theta \,
\nabla_z p \rangle = C_\theta \, \beta \, \langle \theta^2
\rangle$ with the positive dimensionless empirical constant $C_\theta = 1 -\tilde C_\theta$
that is less than 1. We also take into account that $\langle \theta \,
\nabla_i p \rangle$ vanishes, where $i=x, y$.
The justification of these assumptions have been discussed in different contexts. \cite{ZKR07,ZKR13}
}
\end{itemize}

Note that the Kolmogorov hypothesis related to the dissipation rates of the second moments in stably stratified turbulence implies that the normalized dissipation time scale of TPE, $t_\theta/t_{\rm T} \equiv C_{\rm p}$, turbulent heat flux, $t_{\rm F}/t_{\rm T} \equiv C_{\rm F}$, the components $E_\alpha$ of TKE, $t_{\alpha\alpha}/t_{\rm T}$ and off-diagonal $\tau_{iz}$ components of the Reynolds stress, $\tau_{iz}/t_{\rm T} \equiv C_\tau$ are empirical constants. These dissipation time scales are normalised by the dissipation time scale of TKE. Generally, these ratios of the dissipation time scales can be functions of the gradient Richardson number.
For instance, recent direct numerical simulations \cite{KMG20,ZDG19} of a shear produced stably stratified turbulence in Couette flow performed for the gradient Richardson number ${\rm Ri} \leq 0.17$ have shown that these ratios of the dissipation time scales are weakly decreasing functions of the gradient Richardson number.
In these DNS, the Reynolds numbers based on the turbulent velocities and integral time scales
are not larger than $10^3$, while in the atmospheric turbulence the Reynolds numbers are about $10^6$--$10^7$. In addition, the size of the inertial subrange of scales where the Kolmogorov spectrum for the turbulent kinetic energy
has been observed in these simulations, is only one decade.
Since for the gradient Richardson number larger than $0.17$, there are no available information about these ratios of the dissipation time scales, we do not take into account these effects in the present study. The term $\rho_0^{-1} \, \langle \theta \,\nabla_z p \rangle$ in Eq.~(\ref{C3}) can also contribute to the classical gradient term, $\propto E_{z} \, \nabla_z \overline{\Theta}$
in the vertical turbulent flux of potential temperature \cite{KMG20}. But here we neglect this effect as well.

The EFB turbulence closure implies that turbulence is maintained by the velocity shear at any stratification.
\cite{ZKR07,ZKR13} Indeed, the buoyancy flux, $\beta \, F_z$, appears in Eqs.~(\ref{C1}) and~(\ref{X24}) with opposite signs and describes the energy exchange between the densities of the turbulent kinetic energy and turbulent potential energy.
Since in the budget equations~(\ref{C1}) and~(\ref{X24}), the buoyancy fluxes, $\pm \beta \, F_z$ enter with opposite signs, they cancel each other in the budget equation~(\ref{X26}) for the total turbulent energy density.
Therefore, as follows from Eq.~(\ref{X26}), the density of the total turbulent energy is independent of the buoyancy. This implies that there are no grounds to consider the buoyancy-flux term in Eq.~(\ref{C1}) for the turbulent kinetic energy density as an ultimate ''killer" of turbulence.
When the rates of the production and dissipation of the density of the total turbulent energy are compensated, the total turbulent energy is conserved. This implies that an increase of vertical gradient of the mean potential temperature, increases the buoyancy and decreases the density of turbulent kinetic energy, but it increases the turbulent potential energy density, so that the total turbulent energy is conserved.

The main mechanism for the self-regulation of the stably stratified turbulence is as follows. \cite{ZKR07,ZKR13} In a steady state and homogeneous regime of turbulence,
the budget equation~(\ref{C4}) for the vertical turbulent flux $F_z$ of the potential temperature yields:
\begin{eqnarray}
F_z = - C_F \, t_{\rm T} \, \langle u_z^2 \rangle \nabla_z \overline{\Theta} + 2 \, C_\theta \, C_F \, t_{\rm T} \, \beta \, E_\theta .
\label{X27}
\end{eqnarray}
Equation~(\ref{X27}) implies that an increase of the vertical gradient of the mean potential temperature increases the turbulent potential energy $E_{\rm P}$ (and it increases $E_\theta$), but it also decreases
the vertical flux of potential temperature.
This is because two contributions  to the vertical turbulent flux $F_z$ (the classical gradient contribution, $- C_F \, t_{\rm T} \, \langle u_z^2 \rangle \, \nabla_z \overline{\Theta}$, and the non-gradient contribution, $2 \, C_\theta \, C_F \, t_{\rm T} \, \beta \, E_\theta$) have opposite signs. Therefore, this feedback closes a loop, i.e., this effect decreases the buoyancy and maintains the stably stratified turbulence for any gradient Richardson numbers.

Thus, the correct mechanism of self-existence of a stably stratified turbulence includes two steps:
(i) conversion of turbulent kinetic energy into turbulent potential energy with increasing the vertical gradient of the mean potential temperature;
(ii) self-control feedback of the negative, down-gradient turbulent heat transfer through efficient generation of the counteracting, positive, non-gradient heat transfer by turbulent potential energy.
Due to this feedback, the stably stratified turbulence is maintained up to strongly supercritical stratifications. This explains the absence of critical gradient Richardson number as a threshold for existence of stably stratified turbulence. \cite{ZKR07,ZKR13} Actually the ''critical Richardson number", treated many years as a threshold between the turbulent and laminar regimes, separates two turbulent regimes: the strong turbulence typical of atmospheric boundary layers and the weak three-dimensional turbulence typical of the free atmosphere or deep ocean, and characterized by strong decrease in heat transfer in comparison to momentum transfer.

To quantify the stably stratified turbulence, the following basic dimensionless parameters are used:
\begin{itemize}
\item{
the gradient Richardson number,
\begin{eqnarray}
{\rm Ri} = {N^2 \over S^2} ,
\label{C12}
\end{eqnarray}
}
\item{the flux Richardson number,
\begin{eqnarray}
{\rm Ri}_{\rm f} = {- \beta \, F_z \over K_{\rm M} S^2} ,
\label{C13}
\end{eqnarray}
}
\item{the turbulent Prandtl number,
\begin{eqnarray}
{\rm Pr}_{_{\rm T}} = {K_{\rm M} \over K_{\rm H}} ,
\label{C14}
\end{eqnarray}
}
\end{itemize}
where $K_{\rm M}$ is the turbulent viscosity, $ K_{\rm H}$ is the turbulent diffusivity,
and $S^2 = (\nabla_z \meanU_x)^2 + (\nabla_z \meanU_y)^2$ is the squared mean velocity shear.

In the framework of the EFB turbulence closure theory \cite{ZKR07,ZKR13},
we use assumptions outlined at the beginning of section~\ref{sect-IIB} for the budget equations~(\ref{C1})--(\ref{C4}) for the density of TKE $E_{\rm K}$,
the intensity of potential temperature fluctuations $E_\theta$, the vertical turbulent flux $F_z$ of potential temperature, the horizontal turbulent flux $F_i$ of potential temperature, the off-diagonal components of the Reynolds stress $\tau_{iz}$
and the vertical density of TKE $E_z$:
\begin{eqnarray}
0 = - \tau_{i z} \, \nabla_z \meanU_i + \beta \, F_z - {E_{\rm K} \over t_{\rm T}},
 \label{APC1}
\end{eqnarray}
\begin{eqnarray}
0 = - F_z \, \nabla_z \overline{\Theta} - {E_\theta \over C_{\rm p} \, t_{\rm T}},
 \label{APC2}
\end{eqnarray}
\begin{eqnarray}
0 = - 2 E_z \, \nabla_z \overline{\Theta} + 2 C_\theta \, \beta \, E_\theta
 - {F_z \over C_{\rm F} \, t_{\rm T}},
\label{APC3}
\end{eqnarray}
\begin{eqnarray}
0 = - F_z \, \nabla_z \meanU_i - {F_i \over C_{\rm F} \, t_{\rm T}} ,   \quad i=x, y,
\label{APCC3}
\end{eqnarray}
\begin{eqnarray}
0= - 2 E_z \, \nabla_z \meanU_i - {\tau_{iz} \over C_\tau \, t_{\rm T}},   \quad i=x, y,
 \label{APC15}
\end{eqnarray}
\begin{eqnarray}
0 = \beta \, F_z  + {1 \over 2} Q_{zz} - {E_{\rm K} \over 3 t_{\rm T}} .
\label{APC4}
\end{eqnarray}
Equation~(\ref{APC15}) yields expressions
for the turbulent fluxes $\tau_{iz}$ of the momentum and the turbulent
viscosity $K_{\rm M}$:
\begin{eqnarray}
&& \tau_{iz}= - K_{\rm M} \, \nabla_z \meanU_i ,  \quad i=x, y,
\label{CC7}\\
&& K_{\rm M} = 2 C_\tau \, t_{\rm T} \, E_z .
\label{C7}
\end{eqnarray}
Equation~(\ref{APC3}) allows to obtain expression for
the vertical turbulent flux of potential temperature $F_z= - 2 C_{\rm F} \, t_{\rm T} \,
(E_z - C_\theta \, E_{\rm P}) \, \nabla_z \overline{\Theta}$,
where we take into account that $E_\theta = E_{\rm P} \, N^2/\beta^2$
and $N^2 = \beta \, \nabla_z \overline{\Theta}$.
In particular, the expression for the vertical turbulent flux $F_z$ can be rewritten as
\begin{eqnarray}
F_z= - K_{\rm H} \, \nabla_z \overline{\Theta} ,
\label{CCC8}
\end{eqnarray}
where the coefficient of the turbulent diffusion $K_{\rm H}$ reads:
\begin{eqnarray}
K_{\rm H} = 2 C_{\rm F} \, t_{\rm T} \, E_z \, \left(1 - C_\theta \, {E_{\rm P} \over E_z}\right) .
\label{C8}
\end{eqnarray}
By means of Eq.~(\ref{APCC3}) we find the horizontal turbulent flux $F_i$ of potential temperature:
\begin{eqnarray}
F_i = - C_{\rm F} \, t_{\rm T} \, F_z \, \nabla_z \meanU_i , \quad i=x, y .
\label{CC8}
\end{eqnarray}
Since in a stably stratified turbulence the vertical turbulent flux $F_z$ is negative,
the horizontal turbulent flux $F_i$ of potential temperature
is directed along the wind velocity $\meanU_i$, i.e., Eq.~(\ref{CC8}) describes the co-wind
horizontal turbulent flux.

Below we derive expressions for useful dimensionless parameters as universal functions
of the flux Richardson number.
In particular, the obtained expressions for the turbulent
viscosity $K_{\rm M}$ and the turbulent diffusivity $K_{\rm H}$ allow us
to determine the turbulent Prandtl number ${\rm Pr}_{_{\rm T}} = K_{\rm M}/K_{\rm H}$ as
\begin{eqnarray}
{\rm Pr}_{_{\rm T}} = {\rm Pr}_{_{\rm T}}^{(0)} \, \left(1 -  C_\theta \, {E_{\rm P} \over E_z}
\right)^{-1} ,
\label{APCC14}
\end{eqnarray}
where ${\rm Pr}_{_{\rm T}}^{(0)} = C_\tau / C_{\rm F}$ is the turbulent Prandtl number
at ${\rm Ri}={\rm Ri}_{\rm f}=0$, i.e., for a non-stratified turbulence.
Equations~(\ref{C13}) and ~(\ref{APC1}) yields the expression for the density of TKE, $E_{\rm K}=K_{\rm M}\, S^2 \, t_{\rm T} \, (1-{\rm Ri}_{\rm f})$, while
Eq.~(\ref{APC2}) allows us to find the intensity of potential temperature fluctuations
$E_\theta=- C_{\rm p} \, t_{\rm T} \, F_z \, \nabla_z \overline{\Theta}$.
This equation can be rewritten in terms of the density of turbulent potential energy (TPE)
$E_{\rm P} = - \beta \, F_z \, C_{\rm p} \, t_{\rm T}$, so that the ratio
$E_{\rm K}/E_{\rm P}$ reads
\begin{eqnarray}
{E_{\rm K} \over E_{\rm P}} = {1 - {\rm Ri}_{\rm f} \over C_{\rm p} \, {\rm Ri}_{\rm f}} .
\label{APC5}
\end{eqnarray}
By means of Eq.~(\ref{APC5}) we also obtain the densities of TKE and TPE normalized by the density of the total turbulent energy (TTE), $E_{\rm T}=E_{\rm K}+ E_{\rm P}$,
\begin{eqnarray}
{E_{\rm K} \over E_{\rm T}} = {1 - {\rm Ri}_{\rm f} \over 1 - (1 - C_{\rm p}) \, {\rm Ri}_{\rm f}} ,
\label{C5}\\
{E_{\rm P} \over E_{\rm T}} = {C_{\rm p} \, {\rm Ri}_{\rm f} \over 1 - (1 - C_{\rm p}) \, {\rm Ri}_{\rm f}} .
\label{C6}
\end{eqnarray}
Equations~(\ref{APC1}) and~(\ref{C7}) allow us to obtain
the dimensionless ratio
\begin{eqnarray}
\left({\tau \over E_{\rm K}}\right)^2 = {2 C_\tau \, A_z \over 1 - {\rm Ri}_{\rm f}} ,
\label{C9}
\end{eqnarray}
where $A_z \equiv E_z / E_{\rm K}$ is the vertical share of TKE, $\tau = \left(\tau_{xz}^2 + \tau_{yz}^2\right)^{1/2} = K_{\rm M} \, S$ and $\tau_{ij} =\langle u_i \, u_j \rangle$ is the Reynolds stress.
By means of Eqs.~(\ref{C7}) and~(\ref{C9}), we find the expression for another useful dimensionless parameter:
\begin{eqnarray}
\left(S \, t_{\rm T}\right)^2 ={1 \over 2 \, C_\tau  \,  A_z \, (1 - {\rm Ri}_{\rm f})} .
\label{D5}
\end{eqnarray}
In addition, Eqs.~(\ref{C7}) and~(\ref{D5}) allow us to obtain
the dimensionless ratio
\begin{eqnarray}
{\beta \, F_z \, t_{\rm T} \over E_{\rm K}} = - {{\rm Ri}_{\rm f} \over 1 - {\rm Ri}_{\rm f}} ,
\label{DD5}
\end{eqnarray}
while Eqs.~(\ref{APC2}) and~(\ref{CCC8}) yield
the dimensionless ratio
\begin{eqnarray}
&&  {F_z^2 \over E_{\rm K} \, E_\theta} = {2 C_\tau \, A_z \over C_{\rm p}
\, {\rm Pr}_{_{\rm T}}} .
\label{C10}
\end{eqnarray}
Finally, applying Eqs.~(\ref{APCC14}) and~(\ref{APC5}), we arrive at the useful expression
for the turbulent Prandtl number ${\rm Pr}_{_{\rm T}}$:
\begin{eqnarray}
{\rm Pr}_{_{\rm T}}({\rm Ri}_{\rm f}) = {\rm Pr}_{_{\rm T}}^{(0)} \, \left(1 -  {C_\theta \, C_{\rm p} \, {\rm Ri}_{\rm f} \over \left(1 - {\rm Ri}_{\rm f}\right) \, A_z} \right)^{-1} .
\label{CC14}
\end{eqnarray}
Since the turbulent Prandtl number can be rewritten as
${\rm Pr}_{_{\rm T}} \equiv {\rm Ri} / {\rm Ri}_{\rm f}$,
Eq.~(\ref{CC14}) yields the important expression that relates the gradient Richardson number Ri
and the flux Richardson number ${\rm Ri}_{\rm f}$:
\begin{eqnarray}
{\rm Ri}({\rm Ri}_{\rm f}) = {\rm Pr}_{_{\rm T}}^{(0)} \, {\rm Ri}_{\rm f} \, \left(1 -  {C_\theta \, C_{\rm p} \, {\rm Ri}_{\rm f} \over \left(1 - {\rm Ri}_{\rm f}\right) \, A_z} \right)^{-1} .
\label{APC14}
\end{eqnarray}
Expressions~(\ref{C9})--(\ref{D5}) and~(\ref{C10})--(\ref{APC14}) contain the vertical share of TKE, $A_z \equiv E_z / E_{\rm K}$, that will be determined below.

In a shear-produced turbulence, the mean wind shear generates the energy of longitudinal velocity fluctuations $E_x$, which in turns feeds the transverse $E_y$  and the vertical $E_z$ components
of turbulent kinetic energy. The inter-component energy exchange term $Q_{\alpha\alpha}$ in Eq.~(\ref{C4}) is traditionally parameterized through the ''return-to-isotropy" hypothesis (see below). \cite{R51} However, a stratified turbulence is usually anisotropic, and the inter-component energy exchange term $Q_{\alpha\alpha}$ should depend on the flux Richardson number ${\rm Ri}_{\rm f}$.
We adopt another model for the inter-component energy exchange term $Q_{\alpha\alpha}$ which generalizes the ''return-to-isotropy" hypothesis to the case of the stably stratified turbulence.
In this model we use the normalised flux Richardson number ${\rm Ri}_{\rm f}/R_\infty$ varying from 0 for a non-stratified turbulence to 1 for a strongly stratified turbulence,
where the limiting value of the flux Richardson number, ${\rm R}_\infty \equiv
{\rm Ri}_{\rm f}|_{_{{\rm Ri} \to \infty}}$, is defined for very strong stratifications
when the gradient Richardson number ${\rm Ri} \to \infty$.
This model for the inter-component energy exchange term $Q_{\alpha\alpha}$ is described by
\begin{eqnarray}
Q_{xx} = - {2(1 + C_{\rm r}) \over 3 t_{\rm T}} \,\left(3 E_x - E_{\rm int}\right) ,
\label{AP1}
\end{eqnarray}
\begin{eqnarray}
Q_{yy} = - {2(1 + C_{\rm r}) \over 3 t_{\rm T}} \,\left(3 E_y - E_{\rm int}\right) ,
\label{AP2}
\end{eqnarray}
\begin{eqnarray}
Q_{zz} = - {2(1 + C_{\rm r}) \over 3 t_{\rm T}} \,\left(3 E_z - 3E_{\rm K}
+ 2 E_{\rm int}\right) ,
\label{AP3}
\end{eqnarray}
where
\begin{eqnarray}
E_{\rm int} = E_{\rm K} + {{\rm Ri}_{\rm f} \over R_\infty} \, \left({C_{\rm r} \over 1 + C_{\rm r}}\right) \,\left[C_0 \, E_{\rm K} - (1 + C_0) \, E_z\right] ,
\nonumber\\
\label{AP4}
\end{eqnarray}
$C_0$ and $C_{\rm r}$ are the dimensionless
empirical constants. When ${\rm Ri}_{\rm f}=0$,
Eqs.~(\ref{AP1})--(\ref{AP4}) describe the ''return-to-isotropy" hypothesis. \cite{R51}
Thus, by means of  Eqs.~(\ref{APC4}), (\ref{DD5}) and~(\ref{AP3})--(\ref{AP4}), we determine
the vertical share of TKE $A_z \equiv E_z / E_{\rm K}$ as a function
of the flux Richardson number ${\rm Ri}_{\rm f}$:
\begin{eqnarray}
A_z({\rm Ri}_{\rm f})  = {C_{\rm r} \, \left(1 - 2 C_0 \, {\rm Ri}_{\rm f} / {\rm R}_\infty \right) - 3 \, \left({\rm Ri}_{\rm f}^{-1} - 1\right)^{-1} \over 3 + C_{\rm r} \, \left[3 - 2 (1 + C_0) \, {\rm Ri}_{\rm f} / {\rm R}_\infty \right]} .
\nonumber\\
\label{C11}
\end{eqnarray}
According to Eq.~(\ref{C11}), the vertical share  $A_z$ of TKE varies between
$(A_z)_{_{{\rm Ri} \to 0}} \equiv A_z^{(0)} = C_{\rm r}/3(1+C_{\rm r})$ for a non-stratified turbulence and $(A_z)_{_{{\rm Ri} \to \infty}} \equiv A_z^{(\infty)}$ for a strongly stratified turbulence,
where
\begin{eqnarray}
A_z^{(\infty)} ={C_{\rm r}(1 -2 C_0) - 3 \, \left({\rm R}_\infty^{-1} -1\right)^{-1}
\over 3 + C_{\rm r}(1 -2 C_0)} .
\label{C12}
\end{eqnarray}
When there is an isotropy in the horizontal plane, the shares of TKE $A_x \equiv E_x / E_{\rm K}$ and $A_y \equiv E_y / E_{\rm K}$ in horizontal directions are given by
\begin{eqnarray}
A_x=A_y={1 \over 2} \, (1-A_z) .
\label{CC11}
\end{eqnarray}

Now we derive expression for the ratio of the vertical turbulent dissipation length scale $\ell_z=t_{\rm T} \, E_z^{1/2}$ and the local Obukhov length scale $L$ defined as \cite{O46}
\begin{eqnarray}
L= {\tau^{3/2} \over - \beta \, F_z} .
\label{T10}
\end{eqnarray}
To this end we use  Eqs.~(\ref{C13}) and~(\ref{T10}),
which yield
\begin{eqnarray}
K_{\rm M} = {\rm Ri}_{\rm f} \, \tau^{1/2} \, L  .
\label{TT10}
\end{eqnarray}
By means of Eqs.~(\ref{C7}), (\ref{C9}) and~(\ref{TT10}),  we obtain the ratio $\ell_z/L$ as the function of the flux Richardson number:
\begin{eqnarray}
{\ell_z \over L} =\left(2 \, C_\tau \right)^{- 3/4} \, {A_z^{-1/4} \, {\rm Ri}_{\rm f} \over  \left(1 - {\rm Ri}_{\rm f}\right)^{1/4}} .
\label{FF8}
\end{eqnarray}

\begin{figure}
\centering
\includegraphics[width=12cm]{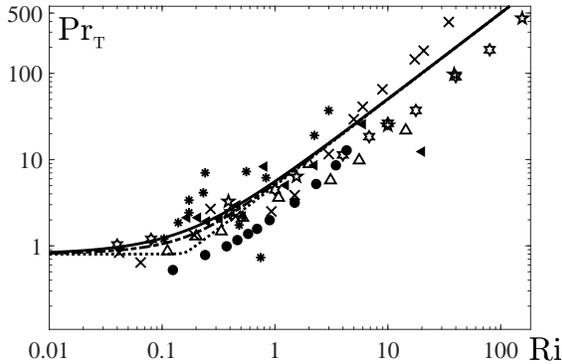}
\caption{\label{Fig1} The turbulent Prandtl number ${\rm Pr}_{_{\rm T}}$ versus
the gradient Richardson number ${\rm Ri}$ for $A_z^{(\infty)}=$ $10^{-3}$ (dotted);  $0.1$ (dashed-dotted); $0.2$ (solid).
Comparison with data of meteorological observations: slanting black triangles \cite{K78}, snowflakes \cite{BB97}; laboratory experiments: slanting crosses \cite{RK04},  six-pointed stars \cite{O01}, black circles \cite{SF01}; DNS: five-pointed stars \cite{SR10}; LES: triangles \cite{ZKR07}.}
\end{figure}

\begin{figure}
\centering
\includegraphics[width=12cm]{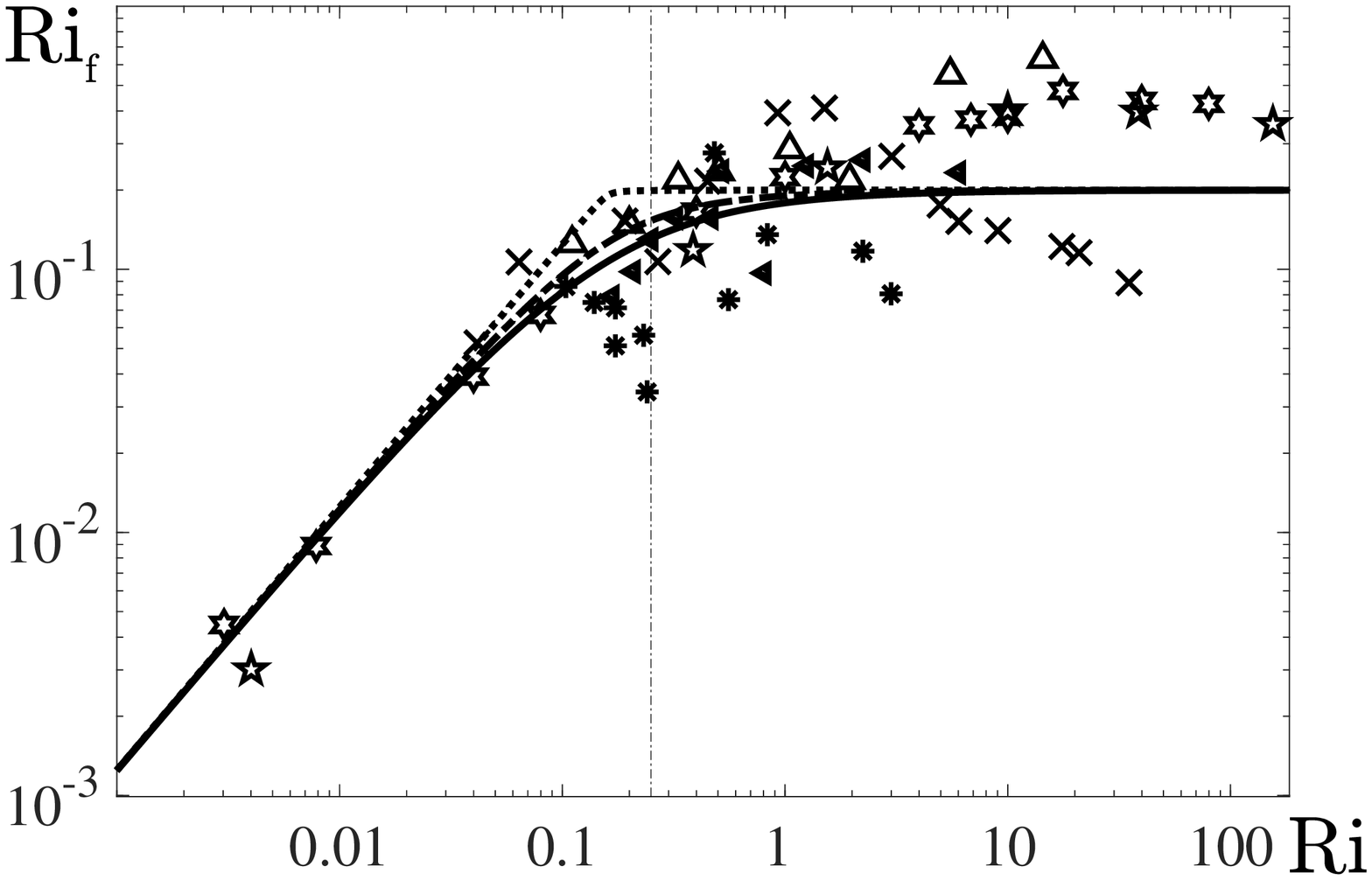}
\caption{\label{Fig2} The flux Richardson number ${\rm Ri}_{\rm f}$ versus
the gradient Richardson number ${\rm Ri}$ for $A_z^{(\infty)}=$ $10^{-3}$ (dotted);  $0.1$ (dashed-dotted); $0.2$ (solid). Comparison with data of meteorological observations:
slanting black triangles \citep{K78}, snowflakes \citep{BB97}; laboratory experiments: slanting crosses \citep{RK04}, six-pointed stars \citep{O01}, black circles \citep{SF01}; DNS: five-pointed stars
\citep{SR10}; LES: triangles \citep{ZKR13}.}
\end{figure}

\begin{figure}
\centering
\includegraphics[width=12cm]{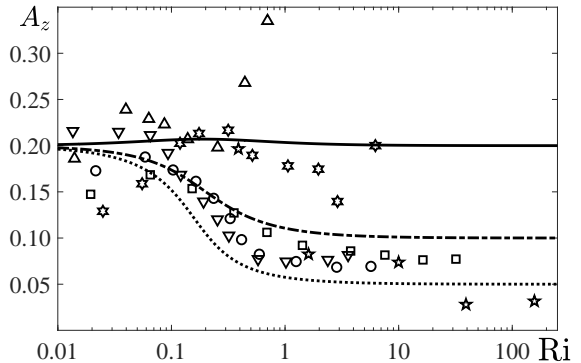}
\caption{\label{Fig3} The vertical share of TKE $A_z \equiv E_z / E_{\rm K}$ versus
the gradient Richardson number ${\rm Ri}$ for $A_z^{(\infty)}=$ $0.05$ (dotted);  $0.1$ (dashed-dotted); $0.2$ (solid). Comparison with data of meteorological observations: squares \citep{MV05}, circles \citep{UC02}, overturned triangles \citep{PC02,BAN02}, six-pointed stars \citep{EA2000}; laboratory experiments: six-pointed stars \citep{O01}; DNS: five-pointed stars \citep{SR10}.}
\end{figure}

\begin{figure}
\centering
\includegraphics[width=12cm]{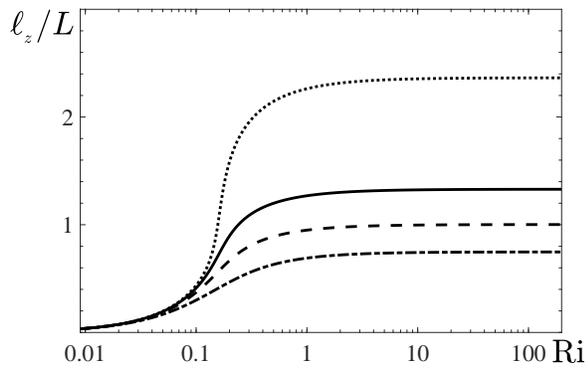}
\caption{\label{Fig4} The normalised vertical turbulent dissipation length scales $\ell_z/L$ versus
the gradient Richardson number ${\rm Ri}$ for $A_z^{(\infty)}=$ $10^{-3}$ (dotted); $3.1 \times 10^{-3}$ (dashed); $10^{-2}$ (solid); $0.1$ (dashed-dotted).}
\end{figure}

For illustration, in Figs.~\ref{Fig1}--\ref{Fig4} we show the dependencies of
the following parameters on the gradient Richardson number ${\rm Ri}$ for different values of the parameter $A_z^{(\infty)}$:
\begin{itemize}
\item{
the turbulent Prandtl number ${\rm Pr}_{_{\rm T}}({\rm Ri})$,
given by Eq.~(\ref{CC14}), see Fig.~\ref{Fig1};
}
\item{
the flux Richardson number ${\rm Ri}_{\rm f}({\rm Ri})$,
given by Eq.~(\ref{APC14}), see Fig.~\ref{Fig2};
}
\item{
the vertical share of TKE $A_z({\rm Ri}) \equiv E_z / E_{\rm K}$,
given by Eq.~(\ref{C11}), see Fig.~\ref{Fig3};
}
\item{
the ratio $\ell_z / L$,
given by Eq.~(\ref{FF8}), see Fig.~\ref{Fig4}.
}
\end{itemize}
The theoretical Ri-dependencies are compared with data of meteorological observations, laboratory experiments, DNS and LES. Figures~\ref{Fig1}--\ref{Fig3} demonstrate reasonable agreement
between theoretical predictions based on the EFB turbulence theory and data obtained from atmospheric and laboratory experiments, LES and DNS.

Data for ${\rm Pr}_{_{\rm T}}({\rm Ri})$ at small gradient Richardson number ${\rm Ri}$ in Fig.~\ref{Fig1} are consistent with the commonly accepted empirical estimate of ${\rm Pr}_{_{\rm T}}^{(0)} = 0.8$ \cite{CH02,FO06,EKR96b}.
The flux Richardson number ${\rm Ri}_{\rm f}$  in the steady-state regime can only increase with the increasing Ri, but obviously cannot exceed unity. Hence it should tend to a finite asymptotic limit (estimated as  ${\rm R}_\infty= 0.2$),  which corresponds to the asymptotically linear Ri-dependence of ${\rm Pr}_{_{\rm T}}$. Thus, the turbulent Prandtl number for strong stratifications is given by
\begin{eqnarray}
{\rm Pr}_{_{\rm T}} = {\rm Pr}_{_{\rm T}}^{(0)} + {{\rm Ri} \over {\rm R}_\infty} .
\label{C16}
\end{eqnarray}
Figure~\ref{Fig2} shows that the flux Richardson number ${\rm Ri}_{\rm f}$ at the gradient Richardson number ${\rm Ri}> 1$ levels off at the limiting value, ${\rm Ri}_{\rm f}={\rm R}_\infty= 0.2$. Figures~\ref{Fig3}--\ref{Fig4} demonstrate that the vertical share of TKE $A_z$ and the ratio $\ell_z / L$ level off at ${\rm Ri}> 1$ as well.

Let us discuss the choice of the dimensionless empirical constants. \cite{ZKR13} There are two well-known universal constants: the limiting value of the flux Richardson number $R_\infty=0.2$ for an extremely strongly stratified turbulence (i.e., for ${\rm Ri} \to \infty$) and the turbulent Prandtl number ${\rm Pr}_{_{T}}^{(0)}=0.8$ for a nonstratified turbulence (i.e., for ${\rm Ri} \to 0$). The  constants $C_{\rm F}=C_\tau/{\rm Pr}_{_{T}}^{(0)}$, where $C_\tau$ is the coefficient determining the turbulent viscosity  ($K_M=2 C_\tau A_z E_K^{1/2} \ell_0$) for a non-stratified turbulence. The constant $C_{\rm p}$ describes the deviation of the dissipation timescale of $E_\theta=\langle \theta^2 \rangle/2$ from the dissipation timescale of TKE. The constants $C_{\rm F}$, $C_{\rm p}$ and $A_z^{(\infty)}$ are determined from numerous meteorological observations, laboratory experiments, direct numerical simulations (DNS) and large eddy simulations (LES). \cite{SF01,O01,SR10,ZKR07,ZKR13,LK16,K78,BB97,RK04,GAF07,NA01,SK00,CM12,MV05,PC02}
The constant $C_\theta$ is given by $C_\theta=\left({\rm R}_\infty^{-1} -1\right) \, A_z^{(\infty)} / C_{\rm p}$ [see Eq.~(\ref{CC14})], and the constant $C_0$ is determined from Eq.~(\ref{C12})
at given $A_z^{(\infty)}$ and $C_{\rm r}$.
We use here the following values of the non-dimensional empirical constants: $C_{\rm F} = 0.125$, $C_{\rm p} = 0.417$, $C_{\rm r} = 3/2$ and $C_\tau=0.1$.
The vertical anisotropy parameter for an extremely strongly stratified turbulence $A_z^{(\infty)}$ is changing in the interval from 0.1 to 0.2 (see Fig.~\ref{Fig3}).

\subsection{Boundary-layer turbulence}

Considering the applications of the obtained results to the atmospheric stably stratified boundary-layer turbulence, we derive below the theoretical relationships potentially useful in modelling applications.
There are two well-known results for the wind shear:
\begin{itemize}
\item{
$S=\tau^{1/2} / \kappa \, z$ at $\varsigma \ll 1$, that yields the log-profile for the mean velocity, and
}
\item{
$S=\tau^{1/2} / {\rm R}_\infty \, L$ when $\varsigma \gg 1$, that follows from Eq.~(\ref{TT10}). Here $\varsigma=\int_0^z \, dz'/L(z')$ is the dimensionless height based on the local Obukhov length scale $L(z)= \tau^{3/2}(z) / [- \beta \, F_z(z)]$, and $\kappa=0.4$ is the von Karman constant.
}
\end{itemize}
The straightforward interpolation between these two asymptotic results for the wind shear,
\begin{eqnarray}
S(\varsigma) = {\tau^{1/2} \over L} \, \left(R_\infty^{-1} +  {1 \over \kappa \, \varsigma} \right),
\label{APPF1}
\end{eqnarray}
yields the vertical profile of the eddy viscosity $K_{\rm M} = \tau/S$ as
\begin{eqnarray}
K_{\rm M}(\varsigma) = \tau^{1/2} \, L \, \, {\kappa \, \varsigma  \over 1 + R_\infty^{-1}
\, \kappa \, \varsigma} .
\label{F1}
\end{eqnarray}
The vertical profile of the flux Richardson number ${\rm Ri}_{\rm f}(z)$ is obtained using  Eqs.~(\ref{TT10}) and~(\ref{F1}):
\begin{eqnarray}
{\rm Ri}_{\rm f}(\varsigma) = {\kappa \, \varsigma \over 1 + R_\infty^{-1} \, \kappa \, \varsigma} .
\label{F2}
\end{eqnarray}
Now we determine the vertical profiles of the turbulent Prandtl number
${\rm Pr}_{_{T}}(z)$ using Eqs.~(\ref{CC14}) and~(\ref{F2}):
\begin{eqnarray}
{\rm Pr}_{_{T}}(\varsigma) ={\rm Pr}_{_{\rm T}}^{(0)} \, \left[1 + {a_1 \, \varsigma + a_2 \, \varsigma^2 \over
1 + a_3 \, \varsigma} \right] ,
\label{F3}
\end{eqnarray}
and the vertical share of TKE $A_z \equiv E_z / E_{\rm K}$
by means of Eqs.~(\ref{C11}) and~(\ref{F2}):
\begin{widetext}
\begin{eqnarray}
A_z(\varsigma) &=& {C_{\rm r} \, {\rm R}_\infty + \kappa \, \varsigma \, \left[C_{\rm r} \, (1 - 2 C_0) - 3 \, ({\rm R}_\infty + \kappa \, \varsigma) \, \left[1 + \kappa \, \varsigma \, \left({\rm R}_\infty^{-1} -1\right) \right]^{-1} \right] \over 3 \,{\rm R}_\infty \, (1 + C_{\rm r}) + \kappa \, \varsigma \, \left[3 + C_{\rm r} \, (1 - 2 C_0)\right]} .
\label{F9}
\end{eqnarray}
\end{widetext}
\noindent
Here ${\rm Pr}_{_{\rm T}}^{(0)} =C_\tau / C_{\rm F}$,
and the coefficients $a_k$ are related to the empirical dimensionless constants:
\begin{eqnarray}
a_1  = 3 \, \kappa \, A_z^{(\infty)} \, \left(1 + C_{\rm r}^{-1}\right) \, \, \left({\rm R}_\infty^{-1} -1\right) ,
\label{F5}
\end{eqnarray}
\begin{eqnarray}
a_2  = {\kappa^2 \, A_z^{(\infty)} \over {\rm R}_\infty} \, \left({\rm R}_\infty^{-1} -1\right)
\, \left(1 - 2 C_0 + 3 C_{\rm r}^{-1}\right) ,
\label{F6}
\end{eqnarray}
\begin{eqnarray}
a_3  &=& \kappa \, \left[2 {\rm R}_\infty^{-1} \, (1 - C_0) - 3 C_{\rm r}^{-1} -1\right] - a_1  .
\label{F7}
\end{eqnarray}
Next, we find the vertical profile of the gradient Richardson number applying
Eqs.~(\ref{APC14}) and~(\ref{F2}):
\begin{eqnarray}
{\rm Ri}(\varsigma) = {\kappa \, \varsigma \, {\rm Pr}_{_{\rm T}}^{(0)} \over 1 + R_\infty^{-1} \, \kappa \, \varsigma } \, \left[1 + {a_1 \, \varsigma + a_2 \, \varsigma^2 \over
1 + a_3 \, \varsigma} \right] .
\label{F4}
\end{eqnarray}
Finally, the vertical profile of the turbulent dissipation length scale $\ell_z(z)$ normalized by the local Obukhov length scale $L(z)$ is obtained by means of Eqs.~(\ref{FF8}) and~(\ref{F2}):
\begin{eqnarray}
{\ell_z \over L} = \left(2 \, C_\tau \right)^{- 3/4} \, {A_z^{-1/4} \, \kappa\, \varsigma \, \left(1 - {\rm Ri}_{\rm f} / {\rm R}_\infty\right) \over \left(1 - {\rm Ri}_{\rm f}\right)^{1/4}} .
\label{F8}
\end{eqnarray}
Equations~(\ref{F1})--(\ref{F8}) for the surface layer ($\varsigma \ll 1$) have been derived in Refs.~\cite{ZKR10,ZKR13}. In the present study we generalize these results for the entire stably stratified boundary layer which are valid for arbitrary values of $\varsigma$.

Equations~(\ref{F1})--(\ref{F8}) are in agreement with the Monin-Obukhov \cite{MO54}
and Nieuwstadt \cite{N84} similarity theories, i.e., the concept of similarity of turbulence in terms of the dimensionless height $\varsigma=\int_0^z \, dz'/L(z')$.
The Monin-Obukhov similarity theory was designed for the ''surface layer" defined as the lower layer
which is 10 \% of the boundary layer, where the turbulent fluxes of momentum $\tau$, heat $F_z$ and
other scalars, as well as the length scale $L$, are approximated by their
surface values. Nieuwstadt (1984) extended the similarity theory to
the entire stably stratified boundary layer employing local $z$-dependent values of the turbulent
fluxes $\tau(z)$ and $F_z(z)$, and the length $L(z)$ instead of their surface values.

\begin{figure}
\centering
\includegraphics[width=11.7cm]{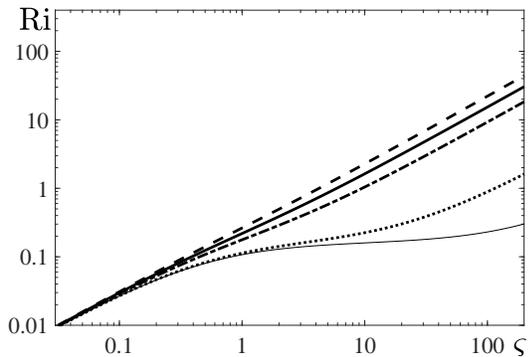}
\caption{\label{Fig5} The gradient Richardson number ${\rm Ri}$ versus $\varsigma=\int_0^z \, dz'/L(z')$ for $A_z^{(\infty)}=$ $10^{-3}$ (solid thin); $10^{-2}$ (dotted);  $0.1$ (dashed-dotted); $0.15$ (solid thick); $0.2$ (dashed).}
\end{figure}

\begin{figure}
\centering
\includegraphics[width=11.7cm]{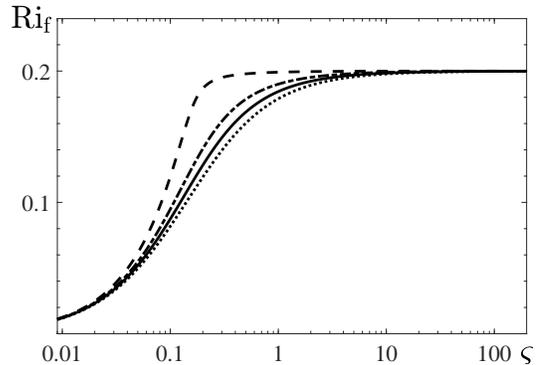}
\caption{\label{Fig6} The flux Richardson number ${\rm Ri}_{\rm f}$ versus $\varsigma=\int_0^z \, dz'/L(z')$ for $A_z^{(\infty)}=$ $10^{-2}$ (dashed);  $0.1$ (dashed-dotted); $0.15$ (solid); $0.2$ (dotted).}
\end{figure}

\begin{figure}
\centering
\includegraphics[width=12cm]{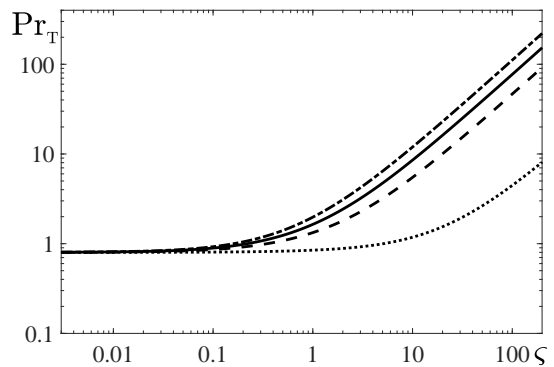}
\caption{\label{Fig7} The turbulent Prandtl number, ${\rm Pr}_{_{\rm T}}$ versus $\varsigma=\int_0^z \, dz'/L(z')$ for $A_z^{(\infty)}=$ $10^{-2}$ (dotted);  $0.1$ (dashed); $0.15$ (solid); $0.2$ (dashed-dotted).}
\end{figure}

\begin{figure}
\centering
\includegraphics[width=12cm]{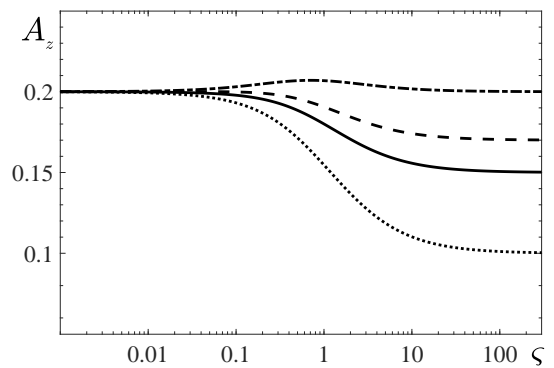}
\caption{\label{Fig8} The vertical share of TKE $A_z \equiv E_z / E_{\rm K}$
versus $\varsigma=\int_0^z \, dz'/L(z')$ for different $A_z^{(\infty)} =$ 0.1 (dotted); 0.15 (solid); 0.17 (dashed) and 0.2 (dashed-dotted).}
\end{figure}

\begin{figure}
\centering
\includegraphics[width=12cm]{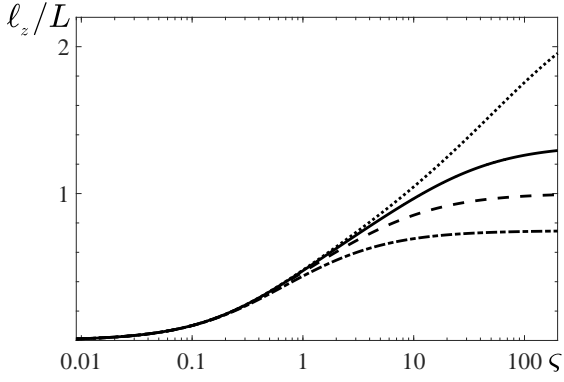}
\caption{\label{Fig9} The normalised vertical turbulent dissipation length scales, $\ell_z/L$ versus $\varsigma=\int_0^z \, dz'/L(z')$ for different $A_z^{(\infty)} =$ 0.1 (dotted); 0.15 (solid); 0.17 (dashed) and 0.2 (dashed-dotted).}
\end{figure}

For illustration, in Figs.~\ref{Fig5}--\ref{Fig9} we plot the vertical profiles of the key turbulent parameters:
\begin{itemize}
\item{
the gradient Richardson number ${\rm Ri}(\varsigma)$, given by Eq.~(\ref{F4}), see Fig.~\ref{Fig5};}
\item{
the flux Richardson number ${\rm Ri}_{\rm f}(\varsigma)$, given by Eq.~(\ref{F2}), see Fig.~\ref{Fig6};
}
\item{
the turbulent Prandtl number ${\rm Pr}_{_{\rm T}}(\varsigma)$, given by Eq.~(\ref{F3}), see Fig.~\ref{Fig7};
}
\item{
the vertical share of TKE $A_z(\varsigma) \equiv E_z / E_{\rm K}$, given by Eq.~(\ref{F9}), see Fig.~\ref{Fig8};
}
\item{
the normalised vertical turbulent dissipation length scales $\ell_z/L$
versus $\varsigma$, given by Eq.~(\ref{F8}), see Fig.~\ref{Fig9}.
}
\end{itemize}
Different lines in Figs.~\ref{Fig5}--\ref{Fig9} correspond to
different values of $A_z^{(\infty)}$ (see below).
As follows from Fig.~\ref{Fig5} for the vertical profile of ${\rm Ri}(z)$,
the gradient Richardson number increases with height,
and the case $\varsigma < 1$ corresponds to Ri $< 0.3$, while
${\rm Ri} > 1$ when $\varsigma> 10$.
Similarly, Fig.~\ref{Fig6} for the vertical profile of ${\rm Ri}_{\rm f}(z)$
shows that the flux Richardson number increases with height
and levels off at $\varsigma > 10$.
The turbulent Prandtl number ${\rm Pr}_{_{\rm T}}$ is constant at $\varsigma < 0.1$
and increases linearly at $\varsigma > 2$ for $A_z^{(\infty)} \geq 0.1$
(and at $\varsigma > 20$ for $A_z^{(\infty)} \ll 0.1$), see Fig.~\ref{Fig7}.
The vertical share of TKE $A_z$ decreases with increasing height
(in all cases except one shown by the dashed-dotted line, see below), and at $\varsigma>100$ it levels off, see Fig.~\ref{Fig8}.
It is clear that the stable stratification, suppressing the
vertical component $E_z$ of TKE, facilitates the energy exchange between the horizontal
velocity energies $E_x$ and $E_y$, and thereby causes a tendency towards isotropy in the horizontal
plane. Equation~(\ref{F8}) for the vertical turbulent dissipation length scale $\ell_z$
has quite expected asymptotes: $\ell_z \propto z$ at $\varsigma \ll 1$ and $\ell_z \propto L$
when $\varsigma >1$ (see Fig.~\ref{Fig9}), where $\varsigma=\int_0^z \, dz'/L(z')$.
In the next sections we develop the energy and flux budget turbulence closure theory for passive scalars.

\section{The EFB theory for passive scalars}

In this section we discuss energy and flux budget turbulence closure theory for passive scalars.

\subsection{Budget equation for turbulent flux of particles}

We consider passive (non-buoyant, non-inertial) particles suspended in the turbulent fluid flow with large Reynolds numbers. Evolution of the particle number density $n_{\rm p}(t, {\bf r})$  (measured in m$^{-3}$) is determined by the following equation:
\begin{eqnarray}
{\partial n_{\rm p} \over \partial t} + ({\bf v}  {\bf \cdot} \bec\nabla) n_{\rm p} = \chi \,\Delta n_{\rm p} ,
\label{D1}
\end{eqnarray}
where ${\bf v}$ is a fluid velocity field and
$\chi$  is the coefficient of molecular (Brownian) diffusion.
Particle number density $n_{\rm p} = \overline{n} + n$ is characterized by the mean value $\overline{n}$ and fluctuations $n$.
Averaging this equation over ensemble of velocity fluctuations, we obtain equation for the mean
particle number density $\overline{n}$. Subtracting this equation from Eq.~(\ref{D1}), we obtain
equation for particle number density fluctuations as
\begin{eqnarray}
{D n \over D t} = - ({\bf u}  {\bf \cdot} \bec\nabla) \overline{n} - ({\bf u}  {\bf \cdot} \bec\nabla) n
+ \langle({\bf u}  {\bf \cdot} \bec\nabla) n \rangle + \chi \,\Delta n .
\label{B2}
\end{eqnarray}
For non-inertial particles, the main effect of turbulent transport is the turbulent diffusion, i.e.,
the turbulent particle flux is $F_i^{({\rm n})} \equiv \langle u_i \, n \rangle = - K_{ij} \, \nabla_j \overline{n}$. This implies that
the quadratic form $K_{ij} \, (\nabla_i \overline{n}) \, (\nabla_j \overline{n})$  should be positively defined. This means that $\partial \overline{n}^2/\partial t < 0$, i.e., the tensor $K_{ij}$ is indeed describes a dissipative process.

Multiplying Eq.~(\ref{B2}) by $u_i$ and the Navier-Stokes equation by $n$, taking the sum and averaging the obtained equation over an ensemble, we obtain the budget equation for the turbulent flux of particles
$F_i^{({\rm n})} = \langle u_i \, n \rangle$:
\begin{eqnarray}
{DF_i^{({\rm n})} \over Dt} + \nabla_z \tilde \Phi_{i}^{({\rm n})} &=& -\langle u_i \, u_j \rangle \, \nabla_j \overline{n}  - F_j^{({\rm n})} \nabla_j \meanU_i + Q_i^{({\rm n})}
\nonumber\\
&& - \varepsilon_i^{({\rm n})} ,
\label{A1}
\end{eqnarray}
where $\tilde \Phi_{i}^{({\rm n})}= \langle u_i \, u_z \, n\rangle + \rho_0^{-1} \, \langle p \, n \rangle \, \delta_{i3}$  is the third-order moment that determines turbulent flux of $F_i^{({\rm n})}$, while $\varepsilon_i^{({\rm n})}= - \chi \, \langle u_i \, \Delta n\rangle - \nu \, \langle n \, \Delta u_i  \rangle$ is the molecular dissipation rate of $F_i^{({\rm n})}$. The Kolmogorov closure hypothesis implies that
$\varepsilon_i^{({\rm n})} = F_i^{({\rm n})} / C_{\rm n} t_{\rm T}$,
where $C_{\rm n}$ is an empirical dimensionless coefficient. The term $Q_i^{({\rm n})} = \rho_0^{-1} \, \langle p \, \nabla_i n \rangle + \beta e_i \, \langle n \, \theta \rangle$ in Eq.~(\ref{A1}) is derived in Appendix A as:
\begin{eqnarray}
Q_i^{({\rm n})} &=& - {C_{\rm D} \over 2}\, \beta \, t_{\rm T} \, e_i \,  F_j^{(n)}  \, \nabla_j \overline{\Theta}
+ \beta \nabla_i \left\langle n \Delta^{-1} \nabla_z \theta\right\rangle ,
\nonumber\\
\label{A5b}
\end{eqnarray}
where ${\bf e}$  is the vertical unit vector, $C_{\rm D}$ is an empirical dimensionless constant. Thus, the budget equation for the turbulent flux of particles can be rewritten as
\begin{eqnarray}
&& {DF_i^{({\rm n})} \over Dt} +  \nabla_z \Phi_{i}^{({\rm n})} =
- {C_{\rm D} \over 2}\, \beta \, t_{\rm T} \, \, e_i \, F_j^{(n)}  \, \nabla_j \overline{\Theta}
-\tau_{ij} \nabla_j \overline{n}
\nonumber\\
&& \quad - F_j^{({\rm n})} \nabla_j \meanU_i
- {F_i^{({\rm n})} \over C_{\rm n} \, t_{\rm T}} ,
\label{A6}
\end{eqnarray}
where $\Phi_{i}^{({\rm n})} = \tilde \Phi_{i}^{({\rm n})} - \beta e_i \left\langle n \Delta^{-1}
\nabla_z \theta\right\rangle$.
Equation~(\ref{A6}) yields the budget equation for the vertical particle flux $F_z^{(n)}$ as
\begin{eqnarray}
&& {DF_z^{({\rm n})} \over Dt} + \nabla_z \Phi_{z}^{({\rm n})} =
- 2 E_z \,  \nabla_z \overline{n}
- {1 \over 2} C_{\rm D} \,  \beta \, t_{\rm T} \, F_z^{({\rm n})} \, \nabla_z \overline{\Theta}
\nonumber\\
&& \quad - {F_z^{({\rm n})} \over C_{\rm n} \, t_{\rm T}} .
\label{A8}
\end{eqnarray}
In Eq.~(\ref{A8}) we have taken into account that $|K_{\rm M} S_i \, \nabla_i \overline{n}| \ll |2 E_z \,  \nabla_z \overline{n}|$, where $S_i \equiv S_{x,y}=\nabla_z \meanU_{x,y}$. We will demonstrate that this condition  provides the positively defined quadratic form $K_{ij} \, (\nabla_i \overline{n}) \, (\nabla_j \overline{n})$.
Equations~(\ref{A6}) and~(\ref{A8}) are complementary equations to the EFB turbulence closure theory discussed in Section~II. \cite{ZKR07,ZKR13}
These equations allow us to determine the turbulent flux of particles $F_i^{({\rm n})}$ at a given gradient of the mean particle number density $\overline{n}$, and the basic turbulent parameters $E_{\rm K}$ and $t_{\rm T}$.

\subsection{Turbulent flux of particles and turbulent diffusion tensor}

In air pollution modelling the particle number density $\overline{n}$ could be strongly heterogeneous in all three directions.
Limiting to the steady-state, homogeneous regime of turbulence, Eq.~(\ref{A8}) reduces to the turbulent diffusion formulation for the vertical turbulent flux of passive scalar (i.e., non-inertial particles):
\begin{eqnarray}
&& F_z^{({\rm n})}= - K_{zz} \, \nabla_z \, \overline{n} ,
\label{D2}
\end{eqnarray}
where
\begin{eqnarray}
K_{zz} &=& K_{\rm M} \, \left({\rm Sc}_{_{\rm T}}^{(0)} + {C_{\rm D} \, {\rm Ri} \over 4 \, A_z \, (1 - {\rm Ri}_{\rm f}) }\right)^{-1} ,
\label{D3}
\end{eqnarray}
${\rm Sc}_{_{\rm T}}^{(0)}=C_\tau / C_{\rm n}$ and $K_{zx}=K_{zy}=0$.
Using Eqs.~(\ref{C7}) and~(\ref{D3}), we determine turbulent Schmidt number as
\begin{eqnarray}
{\rm Sc}_{_{\rm T}} \equiv {K_{\rm M} \over K_{zz}} &=& {\rm Sc}_{_{\rm T}}^{(0)} + {C_{\rm D} \, {\rm Ri} \over 4 \, A_z \, \left(1 - {\rm Ri}_{\rm f}\right)} .
\label{D6}
\end{eqnarray}
In derivation of Eq.~(\ref{D3}), we take into account that
\begin{eqnarray}
N^2 \, t_{\rm T}^2 ={{\rm Ri} \over 2 \, C_\tau  \, A_z \, (1 - {\rm Ri}_{\rm f})} ,
\label{D4}
\end{eqnarray}
which follows from the identities ${\rm Ri}={\rm Ri}_{\rm f} \, {\rm Pr}_{_{\rm T}}$ and
Eq.~(\ref{D5}).

The horizontal components of the turbulent flux of particles can be determined through the steady-state version of Eq.~(\ref{A6}) for a homogeneous stably stratified turbulence:
\begin{eqnarray}
F_x^{({\rm n})}= - K_{xx} \, \nabla_x \,\overline{n} - K_{xz} \, \nabla_z \,\overline{n} ,
\label{D7}\\
F_y^{({\rm n})}= - K_{yy} \, \nabla_y \,\overline{n} - K_{yz} \, \nabla_z \,\overline{n} ,
\label{DD7}
\end{eqnarray}
where
\begin{eqnarray}
K_{xx} &=&  2 \, C_{\rm n} \, t_{\rm T} \, E_x = {A_x \over A_z \, {\rm Sc}_{_{\rm T}}^{(0)}} \, K_{\rm M} ,
\label{D8}
\end{eqnarray}
\begin{eqnarray}
K_{yy} &=&  2 \, C_{\rm n} \, t_{\rm T} \, E_y = {A_y \over A_z \, {\rm Sc}_{_{\rm T}}^{(0)}} \, K_{\rm M} ,
\label{DD8}
\end{eqnarray}
\begin{eqnarray}
K_{xz} &=& -  C_{\rm n} \, t_{\rm T} \, S_x \, K_{zz} ,
\label{D9}
\end{eqnarray}
\begin{eqnarray}
K_{yz} &=& -  C_{\rm n} \, t_{\rm T} \, S_y \, K_{zz} ,
\label{D10}
\end{eqnarray}
where $S_{x,y}=\nabla_z \meanU_{x,y}$ and $K_{xy}=K_{yx}=0$.
In Eqs.~(\ref{D7})--(\ref{DD7}), we have taken into account that $K_{\rm M} \, |S_i \, \nabla_z \overline{n}| \ll |2 E_i \,  \nabla_i \overline{n}|$, where $i=x, y$. We will see that this condition  provides the positively defined quadratic form $K_{ij} \, (\nabla_i \overline{n}) \, (\nabla_j \overline{n})$.
In particular, this condition implies that
\begin{eqnarray}
A_x > {C_{\rm n} \over 4 \, {\rm Sc}_{_{\rm T}}({\rm Ri}_{\rm f})} \, \left[{C_\tau \, A_z({\rm Ri}_{\rm f}) \over 2 (1 -{\rm Ri}_{\rm f})} \right]^{1/2} .
\label{D11}
\end{eqnarray}

Let us discuss the physics related to the off-diagonal terms
of the turbulent diffusion tensor of particles. To this end, we
rewrite the horizontal turbulent flux of particles $F_x^{\rm off} \equiv -  K_{xz} \, \nabla_z \,\overline{n}$ that describes the off-diagonal component $K_{xz}$ of the turbulent diffusion tensor as
\begin{eqnarray}
F_x^{\rm off} = -  C_{\rm n} \, t_{\rm T} \, F_z^{({\rm n})} \nabla_z \meanU_{x} ,
\label{DDD11}
\end{eqnarray}
where $F_z^{({\rm n})}= - K_{zz} \, \nabla_z \, \overline{n}$
is the vertical turbulent flux of particles [see
Eq.~(\ref{D2})].
The turbulent flux of particles $F_x^{\rm off}$ given by
Eq.~(\ref{DDD11}) can be compared with
the horizontal turbulent flux $F_x$ of potential temperature
$F_x = - C_{\rm F} \, t_{\rm T} \, F_z \, \nabla_z \meanU_x$
[see Eq.~(\ref{CC8})]. The latter flux describes the co-wind
horizontal turbulent flux of potential temperature.
For simplicity, we consider here the case when the wind velocity $\meanU_x$
is directed along the $x$-axis.
We remind that in a stably stratified turbulence, the vertical turbulent flux $F_z$
of potential temperature is negative, so that the horizontal turbulent flux $F_x$
of potential temperature is directed along the wind velocity $\meanU_x$.

Contrary, in a convective turbulence the vertical turbulent flux $F_z$
of potential temperature is positive, so that the horizontal turbulent flux $F_x$
of potential temperature is a counter-wind turbulent flux.
The physics of the counter-wind turbulent flux is the following.
Let us consider horizontally homogeneous, sheared convective turbulence.
With increasing height in convection, the mean shear velocity $\meanU_x$ increases
and mean potential temperature $\overline{\Theta}$ decreases. Thus, uprising fluid particles produce positive fluctuations of potential temperature, $\theta>0$
[since $\partial \theta/\partial t \propto - ({\bm u} \cdot
\bec{\nabla}) \overline{\Theta}$], and negative fluctuations of horizontal velocity, $u_x<0$
[since $\partial u_x/\partial t \propto - ({\bm u} \cdot
\bec{\nabla}) \meanU_x$]. This causes negative horizontal temperature flux: $u_x \, \theta<0$. Likewise, sinking fluid particles produce negative fluctuations of potential temperature, $\theta<0$,
and positive fluctuations of horizontal velocity, $u_x>0$, also causing
negative horizontal temperature flux
$u_x \, \theta<0$. This implies that the net horizontal turbulent flux is negative,
$\langle u_x \, \theta\rangle<0$, in spite of a zero horizontal mean temperature gradient.
Thus, the counter-wind turbulent flux of potential temperature
describes modification of the potential-temperature flux by
the non-uniform velocity field. The counter-wind or co-wind turbulent fluxes
are associated with non-gradient turbulence transport of heat.

The comparison of two fluxes, $F_x^{\rm off}$ and $F_x$,
shows that the form of the horizontal turbulent flux of particles $F_x^{\rm off}$
is similar to that of the horizontal turbulent flux of potential temperature $F_x$.
For instance, when the vertical turbulent flux of particles
$F_z^{({\rm n})}$ is positive (or negative), the horizontal turbulent flux of particles $F_x^{\rm off}$
describes the counter-wind (or the co-wind) horizontal turbulent flux of particles.
These turbulent fluxes are associated with non-gradient turbulence transport of particles.

\begin{figure}
\centering
\includegraphics[width=12.0cm]{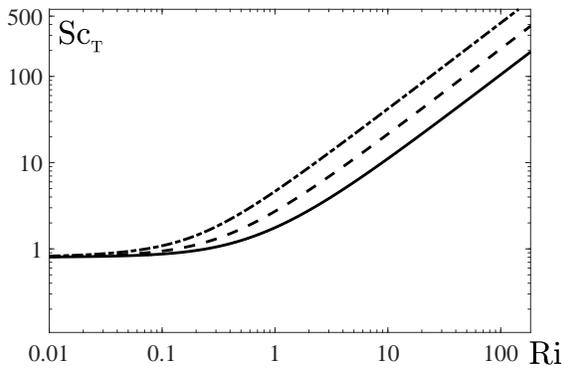}
\caption{\label{Fig10}
The turbulent Schmidt number, ${\rm Sc}_{_{\rm T}}$ versus the gradient Richardson number ${\rm Ri}$
for $A_z^{(\infty)}=0.15$ and different values of $C_{\rm D}$ = 1 (dashed) and 2 (dashed-dotted).
}
\end{figure}

\begin{figure}
\centering
\includegraphics[width=11.8cm]{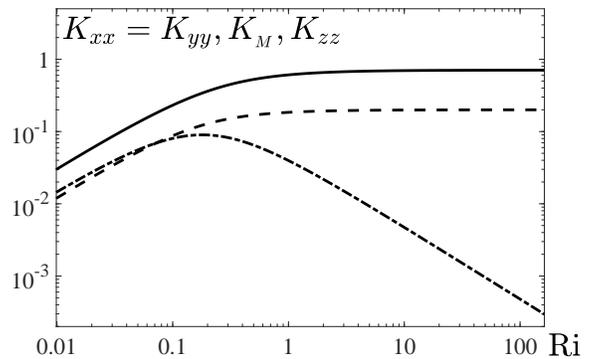}
\caption{\label{Fig11} Diagonal components of the turbulent diffusion tensor: $K_{xx}=K_{yy}$ (solid) and $K_{zz}$ (dashed-dotted), and the eddy viscosity: $K_{\rm M}$ (dashed), normalized by $u_\ast \, L$ versus the gradient Richardson number Ri
for $A_z^{(\infty)}=0.15$.}
\end{figure}

\begin{figure}
\centering
\includegraphics[width=12cm]{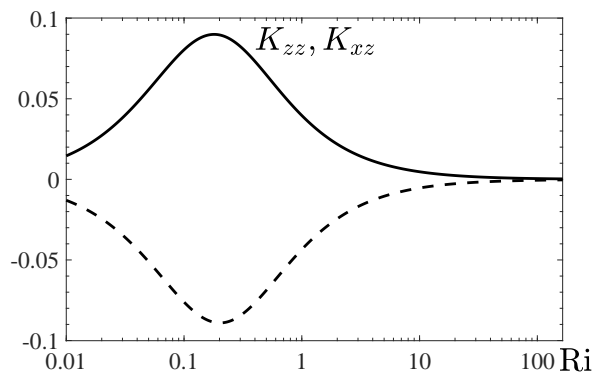}
\caption{\label{Fig12} Off-diagonal component of the turbulent diffusion tensor: $K_{xz}$ (dashed) normalized by $u_\ast \, L$  versus the gradient Richardson number, Ri
for $A_z^{(\infty)}=0.15$. For comparison the diagonal component
$K_{zz}({\rm Ri})$ (solid) is also shown here.}
\end{figure}

For illustration, in Figs.~\ref{Fig10}--\ref{Fig12} we show the dependencies
of the key passive scalar parameters on the gradient Richardson number ${\rm Ri}$:
\begin{itemize}
\item{
the turbulent Schmidt number ${\rm Sc}_{_{\rm T}}({\rm Ri})$, given by Eq.~(\ref{D6}), see Fig.~\ref{Fig10};
}
\item{
the diagonal components $K_{zz}({\rm Ri})$ and $K_{xx}({\rm Ri})=K_{yy}({\rm Ri})$ of the turbulent diffusion tensor, normalized by $u_\ast \, L$ and given by Eqs.~(\ref{D3}) and~(\ref{D8}), see Fig.~\ref{Fig11};
}
\item{
the off-diagonal component $K_{xz}({\rm Ri})$ of the turbulent diffusion tensor, normalized by $u_\ast \, L$ and given by Eq.~(\ref{D9}), see Fig.~\ref{Fig12}.
}
\end{itemize}
Here $u_\ast = \tau^{1/2}$, and we consider for simplicity the case $S_y=0$ and $A_x=A_y$.
The turbulent Schmidt number ${\rm Sc}_{_{\rm T}}({\rm Ri})$  increases linearly with the gradient Richardson number for Ri $> 1$ (see Fig.~\ref{Fig10}).
As follows from Eq.~(\ref{D3}) and Fig.\ref{Fig11}, the vertical turbulent diffusion coefficient $K_{zz}({\rm Ri})$ of particles or gaseous admixtures is strongly suppressed for large gradient Richardson numbers. The vertical turbulent diffusion coefficient and the turbulent Schmidt number behave in the similar fashion as the turbulent temperature diffusion coefficient $K_{\rm H}({\rm Ri})$ and the turbulent Prandtl number ${\rm Pr}_{_{\rm T}}({\rm Ri})$.

The physics of such behaviour of the vertical turbulent diffusion coefficient $K_{zz}({\rm Ri})$ is related to the buoyancy force that causes a correlation between the potential temperature and the particle number density fluctuations $\langle \theta  \, n \rangle$. This correlation is proportional to the product of the vertical
turbulent particle flux $\langle n \, u_z \rangle$ and the vertical gradient of the mean potential temperature $\nabla_z \overline{\Theta}$, i.e., $\langle \theta  \, n \rangle \propto - \langle n \, u_z \rangle \, \nabla_z \overline{\Theta}$. This correlation reduces a standard vertical turbulent particle flux $\langle n \, u_z \rangle$ that is proportional to the vertical gradient of the mean particle number density, $- \nabla_z \overline{n}$.

\begin{figure}
\centering
\includegraphics[width=12cm]{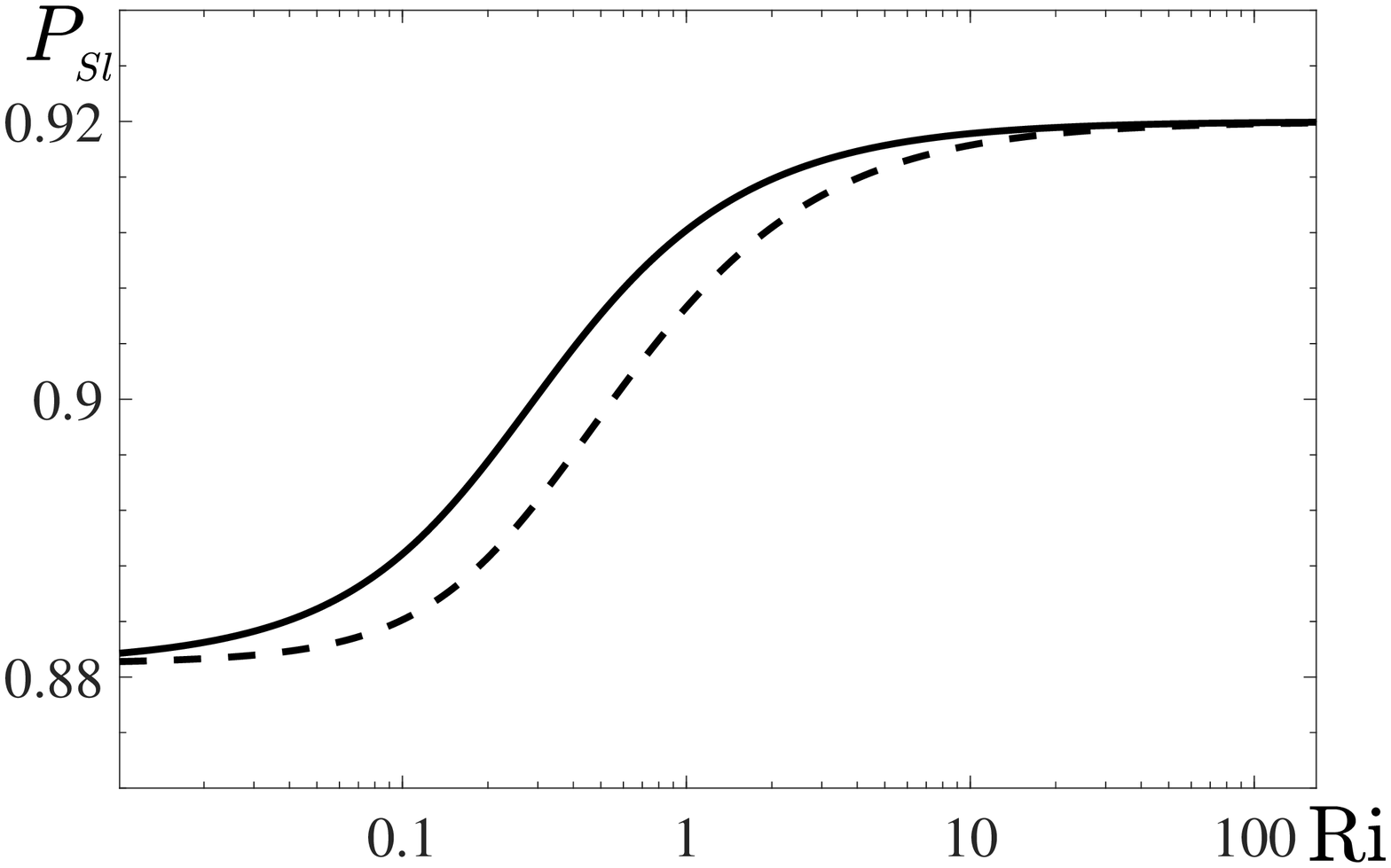}
\caption{\label{Fig13} The parameter $P_{Sl} = 1- K_{xz}^2/(K_{xx}\, K_{zz})$ versus the gradient Richardson number, Ri for $C_{\rm D}$ = 1 (dashed), 2 (dashed-dotted), and for $A_z^{(\infty)}=0.15$.}
\end{figure}

Let us consider for simplicity the case $S_y=0$.
The quadratic form $K_{ij} \, (\nabla_i \overline{n}) \, (\nabla_j \overline{n})$  is positively defined, if the determinant $D_{Sl} = K_{xx}\, K_{yy} \, K_{zz}\, P_{Sl}$ of the symmetric matrix $\tilde K_{ij}$ is positive, where $P_{Sl} = 1 - \tilde K_{xz}^2/(4 K_{xx}\, K_{zz})$ and the diagonal elements $K_{xx}$, $K_{yy}$ and $K_{zz}$ are positive. In the symmetric matrix $\tilde K_{ij}$, the off-diagonal elements $\tilde K_{xz}=\tilde K_{zx}=K_{xz}/2$, and other off-diagonal
elements vanish.
The parameter $P_{\rm Sl}({\rm Ri}) = 1 - K_{xz}^2/(16 K_{xx}\, K_{zz})$
versus the gradient Richardson number is shown in Fig.~\ref{Fig13}, where we use Eqs.~(\ref{D3}), (\ref{D8}) and~(\ref{D9}).
Figure~\ref{Fig13} shows that $P_{\rm Sl}$ and the determinant $D_{\rm Sl}$ are always positive, so that the quadratic form $K_{ij} \, (\nabla_i \overline{n}) \, (\nabla_j \overline{n})$  is positively defined quadratic form.

In view of the above analysis, the down-gradient formulation: $F_\alpha^{({\rm n})}= - K_{\alpha} \, \nabla_\alpha \overline{n}$ , where $K_{\alpha}$  are the turbulent diffusion coefficients along $\alpha=x,y,z$ axes, widely used in operational models, can hardly be considered as satisfactory. It is long ago understood that the linear dependence between the vectors $F_\alpha^{({\rm n})}$  and $\nabla_\alpha \overline{n}$  is characterised by an eddy diffusivity tensor with non-zero off-diagonal terms. \cite{K73} Equations~(\ref{D3}) and~(\ref{D8})--(\ref{D10}) allow determining all components of this tensor.
The equations derived in this section are immediately applicable to turbulent diffusion of gaseous admixtures. In this case $\overline{n}$ and $n$ denote mean value and fluctuations of the admixture concentration (measured in kg/m$^{3}$).

In temperature stratified fluids, there is an additional mechanism of particle transport, namely turbulent thermal diffusion, which causes particle concentration in the vicinity of the mean temperature minimum, i.e., this effect results in the particle transport in the direction opposite to the temperature gradient. \cite{EKR96,EKR97,EKR00,EKR01}
This effect has been detected in laboratory experiments \cite{AEKR17}, DNS \cite{HKRB12} and atmospheric observations. \cite{SEKR09}
In the present paper this mechanism is not considered.

\subsection{Application to boundary-layer turbulence}

\begin{figure}
\centering
\includegraphics[width=12cm]{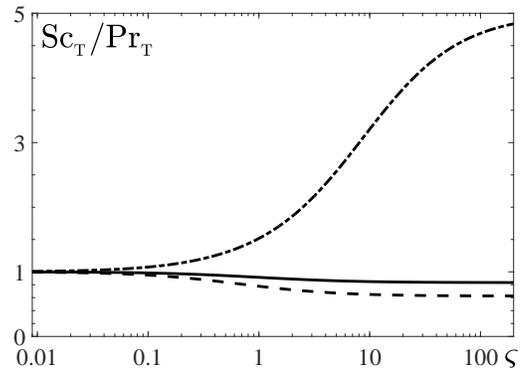}
\caption{\label{Fig14} The ratio ${\rm Sc}_{_{\rm T}}/{\rm Pr}_{_{\rm T}}$ of
turbulent Schmidt number to turbulent Prandtl number versus $\varsigma=\int_0^z \, dz'/L(z')$
for $C_{\rm D}$ = 2 and different values of $A_z^{(\infty)}=$ 0.025 (dashed-dotted), 0.15 (solid); 0.2 (dashed).}
\end{figure}

\begin{figure}
\centering
\includegraphics[width=12cm]{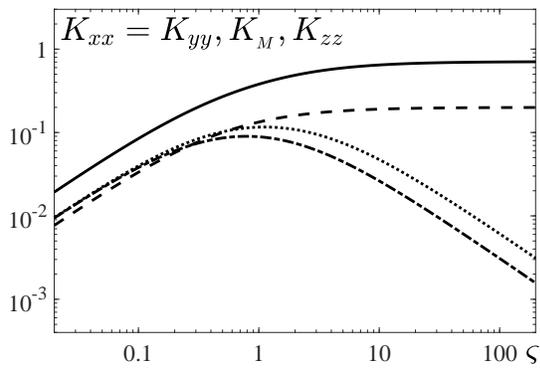}
\caption{\label{Fig15} Diagonal components of the turbulent diffusion tensor: $K_{xx}=K_{yy}$ (solid) and $K_{zz}$ (dashed-dotted), and the eddy viscosity $K_{\rm M}$ (dashed), normalized by $u_\ast \, L$  versus $\varsigma=\int_0^z \, dz'/L(z')$
for $A_z^{(\infty)}=0.15$.}
\end{figure}

\begin{figure}
\centering
\includegraphics[width=12cm]{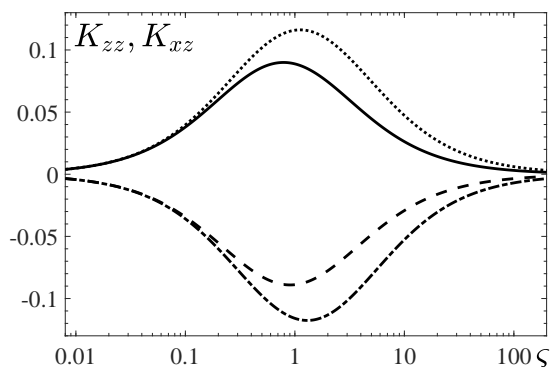}
\caption{\label{Fig16} Off-diagonal component of the turbulent diffusion tensor: $K_{xz}$, normalized by $u_\ast \, L$  versus $\varsigma=\int_0^z \, dz'/L(z')$
for $C_{\rm D}$ = 1 (dashed), 2 (dashed-dotted), and $A_z^{(\infty)}=0.15$.
For comparison the diagonal component $K_{zz}(z/L)$
for $C_{\rm D}$ = 1 (dotted), 2 (solid) is also shown here.}
\end{figure}

Let us consider the applications of the obtained results of particle transport
to the atmospheric stably stratified boundary-layer turbulence.
Equations~(\ref{F1})--(\ref{F4}) allow us to find the vertical profiles of
the turbulent Schmidt number and of the components of the
turbulent diffusion tensor $K_{ij}$ in the atmospheric boundary-layer turbulence.
For illustration, in Figs.~\ref{Fig14}--\ref{Fig16} we plot the vertical profiles of the key
passive scalar parameters:
\begin{itemize}
\item{
the ratio ${\rm Sc}_{_{\rm T}}(\varsigma)/{\rm Pr}_{_{\rm T}}(\varsigma)$ of
turbulent Schmidt number to turbulent Prandtl number, given by Eqs.~(\ref{F3})
and~(\ref{D6}), see Fig.~\ref{Fig14};
}
\item{
the diagonal components of the turbulent diffusion tensor: $K_{zz}(\varsigma)$  and $K_{xx}(\varsigma)=K_{yy}(\varsigma)$, normalized by $u_\ast \, L$ and given by Eqs.~(\ref{D3}) and~(\ref{D8}), see Fig.~\ref{Fig15};
}
\item{
the off-diagonal component of the turbulent diffusion tensor: $K_{xz}(\varsigma)$ normalized by $u_\ast \, L$ and given by Eq.~(\ref{D9}), see Fig.~\ref{Fig16}, where we also use
Eqs.~(\ref{CC11}) and~(\ref{F1})--(\ref{F4}).
}
\end{itemize}
Figure~\ref{Fig14} demonstrates that the ratio ${\rm Sc}_{_{\rm T}}/{\rm Pr}_{_{\rm T}}$ can be more or less than 1 depending on the parameter $C_{\rm D}$.
This implies that the turbulent Schmidt number ${\rm Sc}_{_{\rm T}}$ generally does not coincide with the turbulent Prandtl number ${\rm Pr}_{_{\rm T}}$. This is not surprising, since temperature fluctuations cannot be considered as passive scalar, because they strongly affect velocity fluctuations when the gradient Richardson number is not small. \cite{KLL16}
On the other hand, fluctuations of the number density of non-inertial particles or gaseous admixture behave as passive scalar because they do not affect velocity fluctuations. Only fluctuations of
number density of inertial particles when the mass-loading parameter $m_{\rm p} \, n/ \rho$ is not small ($m_{\rm p} \, n/ \rho > 1$) can affect velocity fluctuations, where $m_{\rm p}$ is a particle mass.

Let us discuss the vertical profiles of the turbulent diffusion tensor $K_{ij}$ shown in Figs.~\ref{Fig15}--\ref{Fig16}. Figure~\ref{Fig15}
demonstrates that the vertical turbulent diffusion coefficient $K_{zz}$ of particles or gaseous admixtures is strongly suppressed when $\varsigma \gg 10$.
Equations~(\ref{D3}), (\ref{D8})--(\ref{D10}) and Figs.~\ref{Fig11}--\ref{Fig12} and~\ref{Fig15}--\ref{Fig16} show that the components $K_{xx}=K_{yy}$ of the turbulent diffusion tensor in the horizontal direction are dominant in comparison with the vertical component $K_{zz}$. This implies that any initially created strongly inhomogeneous distribution of particles (i.e., a strong particle cluster or blob) is evolved into a thin ''pancake" in horizontal plane with very small increase of its thickness in the vertical direction. For instance, when the vertical turbulent heat flux $|F_z| = 0.3$ K m/s, the friction velocity $u_\ast =0.1$ m/s, at the height $z=1$ km, the gradient Richardson number ${\rm Ri}=3$ (see Fig.~\ref{Fig5}), the ratio $K_{xx}/K_{zz} \approx 10^3$ (see Fig.~\ref{Fig15}), so that the horizontal size of the ''pancake" of particles is in 30 times larger than the vertical size.

\section{Conclusions}

We discuss here the energy and flux budget turbulence closure theory for a passive scalar (e.g., non-buoyant and non-inertial particles or gaseous admixtures) in stably stratified turbulence. The EFB turbulence closure theory is based on the budget equations for the turbulent kinetic and potential energies and turbulent fluxes of momentum and buoyancy, and the turbulent flux of particles. The EFB closure theory explains the existence of the shear produced turbulence even for very strong stratifications.

In the framework of the EFB closure theory, we have found that in a steady state homogeneous regime of turbulence, there is a universal flux Richardson number dependence of the turbulent flux of passive scalar described in terms of an anisotropic non-symmetric turbulent diffusion tensor.
We have shown that the diagonal component in the vertical  direction of the turbulent diffusion tensor for particles or gaseous admixtures is strongly suppressed for large gradient Richardson numbers, but the diagonal components in the horizontal directions are not suppressed by strong stratification. We have determined the turbulent Schmidt number defined as the ratio of the eddy viscosity and the vertical turbulent diffusivity of passive scalar which increases linearly with the gradient Richardson number.

We explain these features by the effect of the buoyancy force which causes a correlation between fluctuations of the potential temperature and the particle number density. In particular, this correlation is proportional to the product of the vertical turbulent particle flux and the vertical gradient of the mean potential temperature, which reduces the vertical turbulent particle flux.

In view of applications to the atmospheric stably stratified boundary-layer turbulence, we derive the theoretical relationships for the vertical profiles of the key parameters of stably stratified turbulence measured in the units of the local Obukhov length scale. These relationships
allow us to determine the vertical profiles of the components of the turbulent diffusion tensor
and the turbulent Schmidt number. These results are potentially useful in modelling applications of transport of particles or gaseous admixtures in stably stratified atmospheric boundary-layer turbulence and free atmosphere turbulence.

\begin{acknowledgements}
This paper is dedicated to Prof. Sergej Zilitinkevich (1936-2021) who initiated this work.
This research was supported in part by the PAZY Foundation of the Israel Atomic
Energy Commission (grant No. 122-2020) and
the Israel Ministry of Science and Technology (grant No. 3-16516).
\end{acknowledgements}

\bigskip
\noindent
{\bf DATA AVAILABILITY}
\medskip

Data sharing is not applicable to this article as no new data were created or analyzed in this
study.

\appendix

\section{Derivation of the budget equation for $\langle n \, \theta \rangle$ and Eq.~(\ref{A5b})}

Equations for fluctuations of the potential temperature and the particle number density read:
\begin{eqnarray}
{D\theta \over Dt} &=& - ({\bf u}  {\bf \cdot} \bec\nabla)\, (\overline{\Theta} + \theta) + \langle({\bf u}  {\bf \cdot} \bec\nabla) \theta \rangle + \kappa \, \Delta \theta ,
\label{P1}\\
{D n \over D t} &=& - ({\bf u}  {\bf \cdot} \bec\nabla) (\overline{n} + n)
+ \langle({\bf u}  {\bf \cdot} \bec\nabla) n \rangle + \chi \,\Delta n .
\label{P2}
\end{eqnarray}
Multiplying Eq.~(\ref{P1}) by $n$ and Eq.~(\ref{P2}) by $\theta$, averaging and adding
the obtained equations, we arrive at the budget equation for the correlation function $\langle n \, \theta \rangle$:
\begin{eqnarray}
{D \langle n \, \theta \rangle \over Dt} &+& \nabla_j \Phi_{j}^{({\rm n}\theta)} = - F_j^{({\rm n})} \nabla_j \overline{\Theta} - \varepsilon^{({\rm n}\theta)} .
\label{P3}
\end{eqnarray}
Here $\Phi_{j}^{({\rm n}\theta)}$  is the third-order moment describing the turbulent flux of
the correlation function $\langle n \, \theta \rangle$:
\begin{eqnarray}
\Phi_{i}^{({\rm n}\theta)} = \langle u_i \, n \, \theta \rangle ,
\label{P4}
\end{eqnarray}
and $\varepsilon^{({\rm n}\theta)}$ is the dissipation rate of $\langle n \, \theta \rangle$.
We assume here that the term $- F_j \nabla_j \overline{n}$ contributes to an effective dissipation of
$\langle n \, \theta \rangle$, similarly to the effective dissipation of the Reynolds stress, i.e.,
\begin{eqnarray}
\varepsilon^{({\rm n}\theta)} = - \chi \, \langle \theta \, \Delta n\rangle - \kappa \, \langle n \, \Delta \theta  \rangle - F_j \nabla_j \overline{n} \; .
\label{P5}
\end{eqnarray}
This assumption allows to provide a positive dissipation rate of passive scalar fluctuations.
The effective dissipation rate $\varepsilon^{({\rm n}\theta)}$ can be expressed using the Kolmogorov closure hypothesis:
\begin{eqnarray}
\varepsilon^{({\rm n}\theta)} = {\langle n \, \theta \rangle \over C_{{\rm n}\theta} \, t_{\rm T}} ,
\label{P6}
\end{eqnarray}
where $C_{{\rm n}\theta}$ is the dimensionless constant.
In the steady-state, homogeneous regime of turbulence, Eq.~(\ref{P3}) reduces to the turbulent diffusion formulation:
\begin{eqnarray}
\langle n \, \theta \rangle &=& - C_{{\rm n}\theta} \,  t_{\rm T} \, F_j^{({\rm n})} \nabla_j \overline{\Theta} ,
\label{P7}
\end{eqnarray}
where we consider only gradient approximation neglecting higher spatial derivatives.

Now let us determine the term $Q_i^{({\rm n})} = \rho_0^{-1} \, \langle p \, \nabla_i n \rangle + \beta e_i \, \langle n \, \theta \rangle$. Calculating the divergence of the Navier-Stokes, we obtain
\begin{eqnarray}
\rho_0^{-1} \, \bec{\nabla}^2 p = \beta \, \nabla_z \theta .
\label{P9}
\end{eqnarray}
Applying the inverse Laplacian to Eq.~(\ref{P9}) we arrive at
the following identity:
\begin{eqnarray}
\rho_0^{-1} \, p &=& \beta \, \Delta^{-1} \nabla_z \theta ,
\label{PP8}
\end{eqnarray}
which yields
\begin{eqnarray}
\rho_0^{-1} \, \langle \theta \, \nabla_z p \rangle &=& \beta \, \langle \theta \, \Delta^{-1} \nabla_z^2 \theta \rangle .
\label{P8}
\end{eqnarray}
Using Eqs.~(\ref{PP8}) and~(\ref{P8}), we determine $\rho_0^{-1} \, \langle p \, \nabla_i n \rangle$:
\begin{eqnarray}
\rho_0^{-1} \, \langle p \, \nabla_i n \rangle &=& \beta \, \langle (\nabla_i n) \, \Delta^{-1} \, \nabla_i \theta \rangle = \beta \, \nabla_i \, \langle n \, \Delta^{-1} \, \nabla_z \theta \rangle
\nonumber\\
&& - \beta \, \langle n \, \Delta^{-1} \nabla_z \nabla_i \theta \rangle .
\label{P10}
\end{eqnarray}

Let us determine the correlation function $\langle n \, \Delta^{-1} \nabla_z^2 \theta \rangle$:
\begin{eqnarray}
&& \langle n(t,{\bm x})\, \Delta^{-1} \nabla_z^2 \theta(t,{\bm x}) \rangle =
\lim_{{\bm x} \to {\bm y}} \langle n(t,{\bm x})\Delta^{-1} \nabla_z^2 \theta(t,{\bm y}) \rangle
\nonumber\\
&& \quad \quad = \int \left({k_z^2 \over k^2}\right) \langle n({\bm k}) \theta(-{\bm k}) \rangle \,d{\bm k} .
\label{X31}
\end{eqnarray}
First we consider an isotropic turbulence. The second moment, $\langle n({\bm k}) \theta(-{\bm k}) \rangle$, of potential temperature fluctuations in a homogeneous and incompressible turbulence in a Fourier space reads
\begin{eqnarray}
\langle n({\bm k}) \theta(-{\bm k}) \rangle = {\langle n \, \theta \rangle \, E_{n \theta}(k) \over 4 \pi k^2} ,
\label{X32}
\end{eqnarray}
where the spectrum function is $E_{n \theta}(k) = k_0^{-1} \, (2/3) \, (k / k_{0})^{-5/3}$ for large Reynolds numbers. Here $k_0 \leq k \leq k_{D}$, the wave number $k_{0} = 1 / \ell_0$, the length $\ell_0$ is the integral scale, the wave number $k_{D}=\ell_{D}^{-1}$, and $\ell_{D} = \ell_0 {\rm Pe}^{-3/4}$ is the diffusion scale,
${\rm Pe}=u_0 \, \ell_0/D \gg 1$ is the P\'{e}clet number.
Therefore,
\begin{eqnarray}
&& \langle n(t,{\bm x})\, \Delta^{-1} \nabla_z^2 \theta(t,{\bm x}) \rangle = {\langle n \, \theta \rangle \over 4 \pi} \int_{k_0}^{\infty} E_{n \theta}(k) \,dk
\nonumber\\
&& \quad \quad  \times \int_{0}^{2\pi} \, d\varphi \int_{0}^{\pi} \sin \vartheta \,d\vartheta \,  {k_z^2 \over k^2} ,
\label{XX31}
\end{eqnarray}
where we use the spherical coordinates $(k, \vartheta, \varphi)$ in the ${\bm k}$-space.
For the integration over angles in ${\bm k}$-space we use the following integral:
\begin{eqnarray}
&&\int_{0}^{2\pi} \, d\varphi \int_{0}^{\pi} \sin \vartheta \,d\vartheta \,
{k_i \, k_j \over k^2} = {4 \pi \over 3} \, \delta_{ij} .
\label{X34}
\end{eqnarray}
Therefore, for large P\'{e}clet numbers this correlation function is given by
\begin{eqnarray}
&& \langle n(t,{\bm x})\, \Delta^{-1} \nabla_z^2 \theta(t,{\bm x}) \rangle \approx {1 \over 3} \langle n \, \theta \rangle .
\label{XXX31}
\end{eqnarray}
Now we determine the correlation function $\langle n \, \Delta^{-1} \nabla_z^2 \theta \rangle$ for an anisotropic turbulence:
\begin{eqnarray}
&& \langle n(t,{\bm x})\, \Delta^{-1} \nabla_z^2 \theta(t,{\bm x}) \rangle = \int_{\tilde \ell_z^{-1}}^{\infty} \,dk_z \, \int_{\tilde \ell_h^{-1}}^{\infty} \, k_h \,dk_h \,
\nonumber\\
&& \quad \times \int_0^{2 \pi} \,d\phi \left(1 - {1 \over
1 + k_z^2/k_h^2} \right)\, \langle n({\bm k}) \theta(-{\bm k}) \rangle ,
\label{XX32}
\end{eqnarray}
where we use the cylindrical coordinates $(k_h, \phi, k_z)$ in ${\bm k}$-space,
$\tilde \ell_z$ and $\tilde \ell_h$ are the correlation lengths of the correlation function $\langle n(t, {\bf x}) \, \theta(t, {\bf y}) \rangle$  in the vertical and horizontal directions.
For strongly anisotropic turbulence, i.e., when $\tilde \ell_z \ll \tilde \ell_h$, the contribution of the first term on the right hand side of Eq.~(\ref{XX32}) is dominant, so that
\begin{eqnarray}
&& \langle n(t,{\bm x})\, \Delta^{-1} \nabla_z^2 \theta(t,{\bm x}) \rangle \approx \langle n \, \theta \rangle .
\label{XXX32}
\end{eqnarray}
Therefore, $\langle n(t,{\bm x})\, \Delta^{-1} \nabla_z^2 \theta(t,{\bm x}) \rangle = C_\ast \, \langle n \, \theta \rangle$, where $C_\ast$ varies from 0.3 to 1 depending on the degree of anisotropy of turbulence.

When $i=x,y$, the correlation function $\langle n \, \Delta^{-1} \nabla_z \nabla_i \theta \rangle$
vanishes in an isotropic turbulence.
Equations~(\ref{P7}), (\ref{XXX31}) and~(\ref{XXX32}) yield the expression for the correlation term $Q_i^{({\rm n})}$ as
\begin{eqnarray}
Q_i^{({\rm n})} &=& - {C_{\rm D} \over 2}\, \beta \, t_{\rm T} \, e_i \,  F_j^{(n)}  \, \nabla_j \overline{\Theta}
+ \beta \nabla_i \left\langle n \Delta^{-1} \nabla_z \theta\right\rangle ,
\nonumber\\
\label{AAAA5b}
\end{eqnarray}
where ${\bf e}$  is the vertical unit vector, $C_{\rm D} = C_{{\rm n}\theta} \, (1 + C_\ast)$ is an empirical dimensionless constant.

\end{document}